\documentclass[journal,onecolumn,draftcls]{IEEEtran}
\usepackage{cite}
\usepackage{float}
\usepackage[pdftex]{graphicx}
\usepackage{amssymb}
\usepackage[cmex10]{amsmath}
\usepackage{dsfont}
\usepackage{algorithm}
\usepackage{algpseudocode}%
\usepackage{color}

\DeclareMathOperator*{\argmax}{arg\,max}
\DeclareMathOperator*{\argmin}{arg\,min}
   
\newtheorem{theorem}{Theorem}
\newtheorem{lemma}{Lemma}
\newtheorem{proposition}{Proposition}
\newtheorem{corollary}{Corollary}

\newtheorem{rem}{Remark}

\begin{document}

\title{Increasing Availability in Distributed Storage Systems via Clustering}

\author{Saeid~Sahraei,~\IEEEmembership{Member,~IEEE,}
         and~Michael~Gastpar,~\IEEEmembership{Fellow,~IEEE}
\date{}
\thanks{This work was supported in part by the Swiss National Science Foundation under Grants 169294 and P2ELP2\_178309.}
\thanks{This work was partially presented at the International Symposium on Information Theory (ISIT), Vail, 2018.}
\thanks{S. Sahraei is with the Department of Electrical Engineering, University of Southern California. M. Gastpar is with the school of Computer and Communication Sciences, {\'E}cole Polytechnique F{\'e}d{\'e}rale de Lausanne (emails: ss\_805@usc.edu; michael.gastpar@epfl.ch). }}

\maketitle 

\begin{abstract}
We introduce the Fixed Cluster Repair System (FCRS) as a novel architecture for Distributed Storage Systems (DSS), achieving a small repair bandwidth while guaranteeing a high availability. Specifically we partition the set of servers in a DSS into $s$ clusters and allow a failed server to choose any cluster other than its own as its repair group. Thereby, we guarantee an availability of $s-1$. We characterize the repair bandwidth vs. storage trade-off for the FCRS under functional repair and show that the minimum repair bandwidth can be improved by an asymptotic multiplicative factor of $2/3$ compared to the state of the art coding techniques that guarantee the same availability. We further introduce Cubic Codes designed to minimize the repair bandwidth of the FCRS under the exact repair model. We prove an asymptotic multiplicative improvement of $0.79$ in the minimum repair bandwidth compared to the existing exact repair coding techniques that achieve the same availability. We show that Cubic Codes are information-theoretically optimal for the FCRS with 2 and 3 complete clusters. Furthermore, under the repair-by-transfer model, Cubic Codes are optimal irrespective of the number of clusters.

\end{abstract}
\section{Introduction}

A Distributed Storage System (DSS) is a network consisting of several servers that collectively store a large content. A DSS is designed with two main criteria in mind. Firstly, as individual servers can fail at any given time, the data must be stored in a redundant manner. The objective is to avoid a permanent loss of the data even in the event of multiple simultaneous failures. Secondly, these failed servers must be replaced with new ones efficiently, that is, without generating too much traffic. The abstract model that is commonly used to capture these two aspects is as follows. Suppose we have a file ${\cal M}$ of size $M$ and $n$ servers each with a storage of size $\alpha$. We require that any set of $k$ servers can collectively recover the file ${\cal M}$. This is referred to as the data recovery criterion. Put differently, the network must be resilient to failure of any set of $n-k$ servers. As a server fails, a newcomer must replace it, by connecting to $d$ other servers and downloading $\beta$ units of data from each, thus occupying a repair bandwidth of $\gamma = d\beta$. This is called the repair process and the set of $d$ servers are called the repair group  \cite{dimakis2010network}. 
The definition of repair can be rather ambiguous as there are several different repair models studied in the literature. We are interested in two of them here. ``Exact repair" \cite{rashmi2009explicit,rashmi2011optimal}, where the newcomer must be identical to the failed server; and ``functional repair" \cite{dimakis2010network} where the newcomer has the same functionality as the failed server, meaning that it must be able to participate in future data recovery and repair processes of other servers. These definitions will be made more precise in Section \ref{sec:model}. 

Assuming that a failed server must be able to choose {\it any} set of $d$ servers as its repair group,  the trade-off between $\alpha$ and $\gamma$ has been completely characterized in \cite{dimakis2010network}  under functional repair via a network information flow analysis. Two points on this trade-off are of particular interest: the Minimum Bandwidth Regenerating (MBR) point and the Minimum Storage Regenerating (MSR) point where $\gamma$ and $\alpha$ are minimized, respectively.   As for the exact repair model, explicit codes \cite{rashmi2009explicit,rashmi2011optimal,shah2012interference,cadambe2013asymptotic,tian2015layered,sasidharan2016explicit}  and converse bounds \cite{sasidharan2014improved,prakash2015storage} have been studied in depth. It is known \cite{tian2013rate} that a non-vanishing gap exists between the overall achievable $(\alpha,\gamma)$  region for the two repair models.
 
It was observed in \cite{dimakis2010network} that as the size of the repair group, the parameter $d$,  grows large the required repair bandwidth $\gamma$ can be made smaller for a fixed storage size $\alpha$. Setting $d = n-1$, we achieve the best trade-off between $\alpha$ and $\gamma$. However there are important downfalls to setting $d$ very large. A coding scheme that is designed based on say, $d = n-1$ is not optimal for repairing multiple parallel failures. Furthermore, the servers involved in the repair process of one failed server may not be available to perform other tasks. More specifically, an architecture with large $d$ is not suitable for applications that involve reading hot data \cite{rawat2016locality,tamo2014bounds} where multiple parallel reads of the same block might become necessary. Mainly in light of this latter issue, the parameter {\it availability} is defined in the literature. A server in a DSS is said to have (all-symbol) availability $s-1$ if there are $s-1$ disjoint sets of servers that can serve as its repair group. A DSS has availability $s-1$ if all the servers in the DSS have availability $s-1$. 

This parameter has been largely investigated in the context of Locally Repairable Codes (LRC) \cite{papailiopoulos2014locally,sathiamoorthy2013xoring,tamo2016optimal}, i.e., codes for which the size of the repair group can be made much smaller than $k$. The trade-off between locality (size of the repair group) and availability has been extensively studied \cite{rawat2016locality,tamo2014bounds,pamies2013locally,wang2015achieving,huang2015linear}. Nevertheless, in the context of LRC the parameter repair bandwidth is typically ignored (and sacrificed).  For instance, the achievability results in \cite{rawat2016locality,tamo2014bounds,pamies2013locally,papailiopoulos2014locally} all have a repair bandwidth of {\it at least} $\gamma = rM/k$ where $r$ is the locality of the code. This is easily outperformed by the codes that achieve the MBR point in \cite{dimakis2010network,rashmi2011optimal}.

In this work we introduce the Fixed Cluster Repair System (FCRS) as a novel architecture which aims at achieving a high availability while maintaining a low repair bandwidth. The main idea is to partition the servers into $s$ clusters of equal size, and a final cluster of size $s_0 = n\mod s$. As a server in a cluster fails, we allow it to choose any of the remaining clusters as its repair group (the last cluster is an exception: as it may not contain as many servers as the other clusters, we exempt it from serving as a repair group). This way, we achieve an availability of $s-1$. It is noteworthy that this clustering is not relevant for the data recovery process, meaning that any set of $k$ servers must be able to recover the file, regardless of which cluster they belong to.   

The term Fixed Cluster Repair System has been specifically chosen to contrast with Adjustable Cluster Repair System (ACRS), a general model where the repair groups of two different servers do not necessarily coincide with each other. Studying an ACRS should lead us to answering a general question. Suppose we are given a DSS consisting of $n$ servers that follow the data recovery and repair requirements discussed above, while guaranteeing an availability $s-1$. What is the trade-off between storage $\alpha$ and repair bandwidth $\gamma$ under these constraints? To the best of our knowledge there has not been any literature so far that specifically addresses this question. However, the random linear codes as well as the explicit codes for the seminal work in \cite{dimakis2010network} serve as achievability results for ACRS. In fact, since a server can choose {\it any} subset of $d$ servers as its repair group in \cite{dimakis2010network} where $d\in[k:n-1]$, it is possible to achieve an availability of $s-1$ for any $s -1 \in [\lfloor\frac{n-1}{k}\rfloor]$. The random linear codes and the Cubic Codes designed for FCRS (Sections \ref{sec:functional} and \ref{sec:cubic}) can be viewed as achievability schemes for ACRS too for any availability $s-1\in[\lfloor\frac{n}{k}\rfloor-1]$. While we emphasize that a comparison with the work in \cite{dimakis2010network} is not entirely fair as the parameter availability has not been a driving motive there, we do find it instructive to demonstrate, through a comparative study, how clustering can help with achieving a low repair bandwidth and a high availability.

Our main objective in this paper is to thoroughly analyze the FCRS under both functional and exact repair models. We will follow a network information flow analysis to completely characterize the $\alpha$ vs $\gamma$ trade-off for the FCRS with arbitrary parameters. An interesting observation is made here. We show that the only adverse affect of increasing the number of clusters in the FCRS is the inevitable decrease in the size of the repair groups. In other words, two FCRSs with $n = ds$ and $n' = ds'$ with respectively $s$ and $s'$ clusters have exactly the same performance in terms of the achievable $(\alpha,\gamma)$ region. By characterizing the entire $(\alpha,\gamma)$ region, we show that for small values of $\gamma$, FCRS performs  better than  \cite{dimakis2010network}. Whereas, on the other end of the spectrum, when $\alpha$ is small, \cite{dimakis2010network} is superior. The improvements offered by the FCRS are most visible at the MBR point itself (the point where the repair bandwidth is minimized), at which we prove an asymptotic multiplicative improvement of $\frac{2}{3}$ over the repair bandwidth compared to \cite{dimakis2010network}, as $s$, $k$ and $n$ grow large.  

Our second contribution is to propose Cubic Codes for the FCRS which are designed to minimize the repair bandwidth under exact repair. Cubic Codes are examples of Fractional Repetition Codes \cite{el2010fractional}, codes that do not require any computation to perform the repair process. More specifically, they are subclasses of Affine Resolvable Designs and generalizations of Grid Codes both discussed in \cite{olmez2016fractional}. When the number of clusters is small (two or three complete clusters with no residual servers), we prove that Cubic Codes do minimize the repair bandwidth for the FCRS. While we do not generalize this  proof of optimality of Cubic Codes to an arbitrary number of clusters, we prove that they achieve an asymptotic (again, as $s$, $k$ and $n$ grow large) multiplicative improvement of  $0.79$ over the repair bandwidth compared to the MBR codes for \cite{dimakis2010network}. Furthermore, we can show that under the slightly more restrictive notion of repair-by-transfer\cite{shah2012distributed}, where no computations are permitted by the newcomer to perform the repair process, Cubic Codes are optimal irrespective of the number of clusters and even if there are residual servers. 

\begin{rem}
Intuitively, there are two properties that distinguish FCRS from the model in \cite{dimakis2010network}. Firstly, since the servers within the same cluster as the failed server cannot help with the repair process, the size of the repair groups is in general smaller. This acts as a disadvantage for FCRS, since the size of the repair group plays a central role in decreasing the repair bandwidth \cite{dimakis2010network}. Secondly, and on the positive side, FCRS is less restricted than the model in \cite{dimakis2010network}, as only certain subsets of the servers (the clusters) must be able to serve as the repair group. As we will see in the following sections, at the MBR point the second factor triumphs and FCRS achieves a lower repair bandwidth. At the MSR point however, the first factor seems to play a more important role (see Section \ref{sec:functionalcomparison} for a comparison).
\end{rem}
\begin{rem}
The improvement that FCRS offers for the repair bandwidth is most pronounced when the number of clusters is large. However, since $\frac{k}{n}\le \frac{1}{s}$, very large number of clusters is of little practical interest. It is therefore important to emphasize that the asymptotic improvements of $0.79$ and $\frac{2}{3}$  for the exact and functional repair models are mostly of theoretical value, as a step towards characterizing the fundamental trade-off between repair bandwidth and availability. Having said this, the analysis in Sections \ref{sec:functionalcomparison} and \ref{sec:cubiccomparison} shows that FCRS improves the state of the art repair bandwidth for values of $s$ as small as 3 (albeit by less than the asymptotic factors), which can be of practical interest.
\end{rem}

Before we move on, it is worth noting that clustering is not a new term or technique in the analysis of Distributed Storage Systems. ``Clustered Storage Systems" have been studied in a series of works \cite{prakash2016generalization,prakash2017storage} where different repair bandwidths are associated to inter-cluster and intra-cluster repair. These models have close connections with the ``rack model" \cite{gaston2013realistic,pernas2013non,tebbi2014code} and are generally motivated by the physical architecture of the network and the fact that the cables/channels which connect the servers within one cluster or rack have higher capacities than the inter-cluster counterparts, which creates the motivation to mostly confine the repair process to within the same cluster as the failed server. They also have slightly different data recovery requirements in \cite{prakash2016generalization}  than \cite{dimakis2010network} and the model studied here. These physical considerations do not play any role in our analysis. We simply assume a completely symmetric structure where the communication bandwidth between any pair of servers is identical.

The rest of the paper is organized as follows. In Section \ref{sec:model} we provide a precise description of FCRS. In Section \ref{sec:functional} we analyze the FCRS with arbitrary parameters under the functional repair model and make a numerical as well as analytical comparison with the results in \cite{dimakis2010network}. In Section \ref{sec:cubic} we introduce Cubic Codes as explicit constructions targeted to minimize the repair bandwidth for the FCRS under exact repair. Comparisons with MBR codes for \cite{dimakis2010network} will follow. In Section \ref{sec:converse}, we provide two converse bounds, respectively indicating that Cubic Codes are optimal for the FCRS under exact repair with $s\le 3$ complete clusters, and under repair-by-transfer with arbitrary parameters. We will conclude the paper in Section \ref{sec:conclusion}.

\section{Model Description}
\label{sec:model}

The FCRS is defined by three parameters $n$, $k$ and $s$. Suppose the network consists of $n = ds + s_0$ servers where $d = \lfloor\frac{n}{s}\rfloor$, $s_0 = n \mod s$, and $2\le s\le\lfloor\frac{n}{k}\rfloor$. We partition these servers into $s+1$ clusters, $s$ of which are of size $d$ and the last of size $s_0$. We have a file ${\cal M}$ of size $H({\cal M}) = M$. Each server $i\in[n]$ is equipped with a memory. We model each memory with a random variable where the server can store a function of ${\cal M}$. Specifically the random variable $X^{(i)}_{j,t}$ represents the content of the $j$'th server in the $i$'th cluster at time-slot $t$ where $(i,j)\in \{[s]\times [d]\} \cup \{\{s+1\}\times[s_0]\}$ and $t\in\mathbb{Z}^+\cup\{0\}$. The purpose of introducing this time parameter is to sort the events by the order at which they occur. We restrict the size of each memory to be bounded by $\alpha$, that is, $H(X^{(i)}_{j,t}) \le \alpha \; \forall  \; i,j,t$. 
 For a set $E\subseteq [d]$ we define $X^{(i)}_{E,t} = \{X^{(i)}_{e,t} \mbox{ s.t. } e\in E\}$. The initial contents of the servers at $t = 0$ must be chosen in such a way that any set of $k$ servers can collectively decode the file ${\cal M}$, irrespective of their clusters. In other words, for any $(E_1,\dots, E_s,E_{s+1})$ that satisfy $E_i\subseteq [d]$ for $i\in[s]$ and $E_{s+1}\subseteq[s_0]$, and $\sum_{i=1}^{s+1} |E_i| = k$, we must have

\begin{figure}[h]
\centering
\includegraphics[scale=0.4]{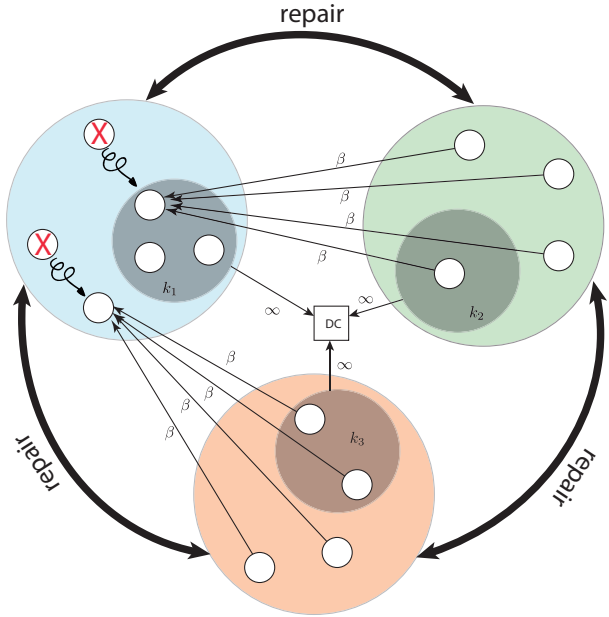}
\caption{The  FCRS with three complete clusters. Two nodes in the blue cluster fail which are repaired by the red and green clusters respectively. A data collector (DC) is connected to $k = k_1 + k_2 + k_3$ servers, one of which is newcomers.}
\label{fig:somemodel}
\end{figure}
\begin{eqnarray}
H({\cal M}|X^{(1)}_{E_1,0},\dots, X^{(s+1)}_{E_{s+1},0}) = 0.
\label{eqn:datarecoveryspecial}
\end{eqnarray}
The servers in the network are subject to failure. To keep track of the order of the events, we assume that time is slotted and that at the end of each time-slot exactly one server fails. Suppose at the end of time-slot $t$, the $\ell$'th server in the $r$'th cluster fails. At the beginning of the next time-slot this server is replaced by a newcomer. 
 A second cluster $i$ will be chosen {\it arbitrarily} such that $i\neq r$. We refer to this as the repair group. The $j$'th server in the repair group transmits $Y^{(r,i)}_{\ell,j,t}$ , a function of $X^{(i)}_{j,t}$ to the newcomer. We limit the size of this message to satisfy $H(Y^{(r,i)}_{\ell,j,t})\le \beta$. Upon receiving $Y^{(r,i)}_{\ell,[d],t}$ the newcomer computes $X^{(r)}_{\ell,t+1}$ as a function of $Y^{(r,i)}_{\ell,[d],t}$. Therefore,  $H(X^{(r)}_{\ell,t+1}| Y^{(r,i)}_{\ell,[d],t}) = 0$. We refer to this process as one round of failure and repair. Due to this requirement, we can assume without loss of generality that $\alpha \le d\beta$. Note that if $(i,j)\neq (\ell,r)$ then $X^{(i)}_{j,t+1} = X^{(i)}_{j,t}$.  In other words, apart from the failed server, the remaining servers remain unchanged at time-slot $t$. 
\begin{rem} More generally, one can assume that several servers fail before any of them is repaired. Such an assumption can be justified in settings where there is no efficient mechanism for prompt detection of failures. However the case of multiple failures can be analyzed in much the same way by ``expanding" the timeline and imposing restrictions on which clusters can participate in the repair process. Specifically, assume that at the end of time-slot $t_0$, $\mu$ servers fail. Let us say all these servers belong to a set of clusters $\Theta$ where $|\Theta|< s$. We can model this as $\mu$ consecutive failures and repairs occurring within time-slots $[t_0: t_0 + \mu -1]$. At any time-slot $t\in [t_0: t_0 + \mu -1]$ none of the clusters containing the servers that have not been repaired yet can participate in the repair process. Even more generally, one may assume that at any given time-slot $t$ a certain number of clusters are unavailable and cannot take part in the repair process (due to being impaired or busy with other tasks). Naturally, if at a given time-slot $t$, $q$ clusters are unavailable (including the cluster that contains the server being repaired), we will only achieve an availability of $s -  q$. Apart from this inevitable inconvenience, our analysis will remain intact and the  repair bandwidth is not worsened due to this assumption (See Section \ref{sec:functional}).
 \end{rem}
 
We will study two different repair models, both of which have been widely studied in the literature \cite{dimakis2010network,ye2017explicit,rashmi2009explicit,rashmi2011optimal,shah2012interference,cadambe2013asymptotic,sasidharan2014improved,prakash2015storage,olmez2016fractional,dimakis2011survey,kamath2013explicit}. 
\begin{itemize}
\item {\it Functional repair}: Under the functional repair model the newcomer may not be identical to the failed server but it must satisfy the data recovery criterion. That is, for any $(E_1,\dots, E_{s+1})$ that satisfy $E_i\subseteq [d]$ for $i\in[s]$ and $E_{s+1}\subseteq[s_0]$ and $\sum_{i=1}^{s+1} |E_i| = k$, we must have
\begin{eqnarray}
H({\cal M}|X^{(1)}_{E_1,t},\dots, X^{(s)}_{E_{s+1},t}) = 0,  \; \forall  \; t\ge 0 .
\label{eqn:datarecoverygeneral}
\end{eqnarray}
\item {\it Exact repair}: Under the exact repair model the content of the newcomer must be identical to the failed server. Therefore, we must have
\begin{eqnarray*}
X^{(i)}_{j,t} = X^{(i)}_{j,t+1} \; \forall \;t,i,j.
\end{eqnarray*}
while studying this model we may omit the subscript $t$ for simplicity and write $X^{(i)}_{j,t}  = X^{(i)}_{j} $. Note that the data recovery criterion automatically holds for the exact repair model, due to \eqref{eqn:datarecoveryspecial}.
\begin{itemize}
\item {\it Repair-by-Transfer:} We will also briefly look at repair-by-transfer \cite{shah2012distributed} which is a sub-model of exact repair. This model requires a newcomer to perform the repair process without any computations. In other words, if the $\ell$'th server in the $r$'th cluster fails, and cluster $i$ is chosen as the repair group, we must have $X^{(r)}_\ell = Y^{(r,i)}_{\ell,[d]}$. In our analysis, we will consider a broader notion of repair-by-transfer which only requires $H( Y^{(r,i)}_{\ell,[d]} |X^{(r)}_\ell) = 0$. 
\end{itemize}


\end{itemize}
The model described above is what we refer to as Fixed Cluster Repair System (FCRS). See Figure \ref{fig:somemodel} for an illustration of an FCRS with three complete clusters and no residual servers ($s_0 = 0$). By contrast, a general DSS (what we referred to as ACRS in the introduction) lacks many of these constraints. A DSS with parameters $(n,k)$ consists of $n$ servers $\{X_1,\dots,X_n\}$ such that any $k$ servers can recover the file ${\cal M}$. A DSS is said to have availability $s-1$ if for each server $X_i$ there are $s-1$ disjoint sets of servers of respective sizes $d^{(i)}_1,\dots,d^{(i)}_{s-1}$ that can serve as its repair group while generating repair bandwidths $d^{(i)}_1\beta^{(i)}_1, \dots , d^{(i)}_{s-1}\beta^{(i)}_{s-1}$, respectively. The repair process can be defined either as functional or exact repair. The repair bandwidth is defined as 
\begin{eqnarray}
\gamma = \max_{i\in[n],j\in[s-1]}d^{(i)}_j\beta^{(i)}_j.
\label{eqn:gammaacrs}
\end{eqnarray}

\section{The Functional Repair Model}
\label{sec:functional}
\begin{figure}[h]
\centering
\includegraphics[scale=0.7]{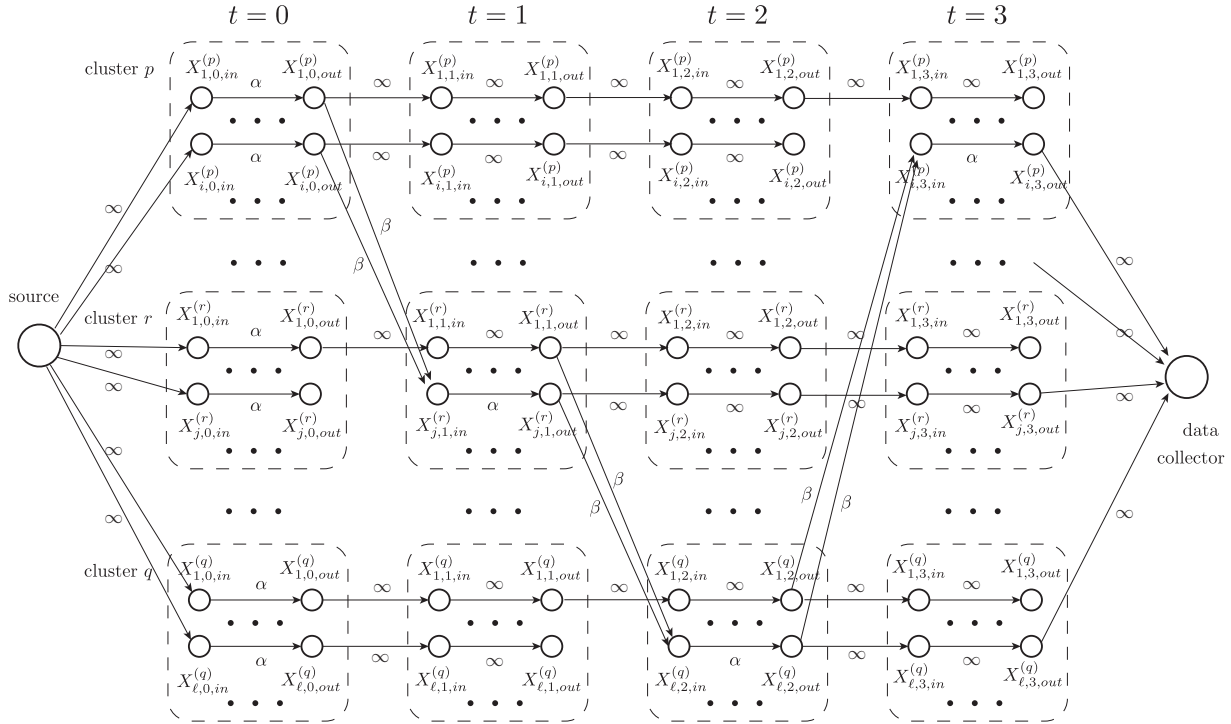} 
\caption{The network information flow model for the FCRS. At the end of $t = 0,1,2$ the servers $X^{(r)}_{j,0}, X^{(q)}_{\ell,1}$ and $X^{(p)}_{i,2}$  fail respectively. These servers are repaired by connecting to clusters $p,r$ and $q$, in that order. Finally a data collector connects to $k$ servers at time-slot $t = 3$ for recovering the file ${\cal M}$. Note that some of these $k$ servers are newcomers.}
\label{fig:evolution3failures}
\end{figure}

In this section, we present a network information flow analysis for the FCRS. Each server $X^{(i)}_{j,t}$ is modeled by a pair of nodes $X^{(i)}_{j,t,in}$ and $X^{(i)}_{j,t,out}$ that are connected with an edge. The sources is directly connected to each node $X^{(i)}_{j,0,in}$ with edges of infinite capacity. Each node $X^{(i)}_{j,0,in}$ is in turn connected to $X^{(i)}_{j,0,out}$ with an edge of capacity $\alpha$. Suppose at the end of time-slot $t-1$ (where $t\ge 1$) the $j$'th server from the $i$'th cluster fails. Assume the newcomer replacing this server is repaired by connecting to the $r$'th cluster. We represent this by $d$ edges which connect $X^{(r)}_{[d],t-1,out}$ to $X^{(i)}_{j,t,in}$. Each of these edges has a capacity of $\beta$. Furthermore, there will be an edge of capacity $\alpha$ from $X^{(i)}_{j,t,in}$ to $X^{(i)}_{j,t,out}$. Since we have only one failure per time-slot, for all other $(i',j')\neq(i,j)$ there will be edges of infinite capacity from $X^{(i')}_{j',t-1,out}$ to $X^{(i')}_{j',t,in}$ and from $X^{(i')}_{j',t,in}$ to $X^{(i')}_{j',t,out}$. At any given time a data collector can be connected to the out nodes of any set of servers of size $k$ with edges of infinite capacity. An illustration has been provided in Figure \ref{fig:evolution3failures} which involves three clusters.  

Our goal in this section is to find the minimum cut that separates any data collector from the source in this graph under all possible failure and repair patterns. As we shall see this minimum cut helps us to characterize the smallest possible value of $\alpha$ for any choice of $\gamma = d\beta$, such that any data collector can recover the file. Furthermore, the trade-off $(\alpha,\gamma)$ characterized by this min-cut is achievable, for instance if we resort to random linear codes \cite{ho2006random}. 

\begin{figure}[H]
\centering
\includegraphics[scale=0.5]{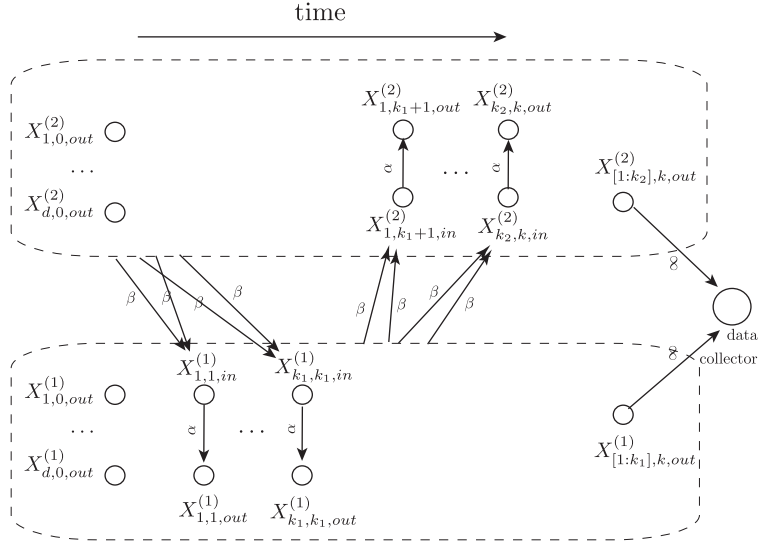}
\caption{Network information flow graph corresponding to a sequence of failures and repairs where $k_1\ge \lceil\frac{k}{2}\rceil$ servers from the first cluster fail, followed by $k_2 = k -k_1$ failures from the second cluster. Each failure in cluster $1$ is repaired by cluster $2$ and vice versa. A data collector is connected to the $k$ newcomers.}
\label{fig:twin}
\end{figure}

Consider a sequence of failures and repairs as depicted in Figure \ref{fig:twin}.  Note that only two clusters participate in this sequence. First, $k_1\ge \lceil\frac{k}{2}\rceil $ servers from the first cluster fail. All of these servers are repaired by connecting to the second cluster. Next, $k_2 = k - k_1$ servers from the second cluster fail. These servers are repaired by the first cluster. Assume a data collector connects to these $k$ newcomers in order to  recover the file ${\cal M}$. As we shall see soon, a simple cut-set argument shows that we must have
\begin{eqnarray*}
M&\le& k_1\alpha + (d-k_1)(k-k_1)\beta.
\end{eqnarray*}
Our first objective is to prove that for any choice of the parameters $\alpha$ and $d\beta$, there exists a $k_1\in [\lceil\frac{k}{2}\rceil:k]$ such that this is the smallest cut which separates any data collector from the source. Let us assume that at some arbitrary point in time, $t_0$, a data collector is connected to $k$ servers which we call $Z_{1,t_0},\dots,Z_{k,t_0}$. For any $i\in[k]$ let $t_i\le t_0$ be the smallest integer such that an edge of infinite capacity exists from $Z_{i,t',out}$ to $Z_{i,t'+1,in}$ for all $t'\in[t_i:t_0]$. If no such $t_i$ exists, set $t_i = t_0$. We say that there is a path from $Z_{i,t_i}$ to $Z_{j,t_j}$ if there exists a $t'\in[t_i:t_j -1]$ such that there is an edge of capacity $\beta$ connecting $Z_{i,t',out}$ to $Z_{j,t'+1,in}$. We can order these $k$ servers such that $i<j$ implies $t_i\le t_j$. As a result, $i < j$ implies there is no path from $Z_{j,t_j}$ to $Z_{i,t_i}$. Let us assume that such an ordering is in place. Define $e_i\in[s+1]$ as the index of the cluster to which $Z_{i,t_0}$ belongs and let
\begin{eqnarray}
c(i,j) = \left|\{\ell \;\; s.t. \;\; \ell\le j \mbox{ and } e_\ell = i\}\right| \mbox{ for } j\in [k],\; i \in [s+1]
\label{eqn:commuli}
\end{eqnarray}
be the number of servers in $Z_{[j],t_0}$ which belong to the $i$'th cluster. Let $F(Z_{[k],t_0})$ be the value of the minimum cut that separates a data collector connecting to $Z_{[k],t_0,out}$ from the source. In order to find this cut, we must decide for any $j\in[k]$ whether to include both $Z_{j,t_j,in}$ and $Z_{j,t_j,out}$ on the sink (data collector) side, or to include $Z_{j,t_j,in}$ on the source side and $Z_{j,t_j,out}$ on the sink side (if we include both $Z_{j,t_j,in}$ and $Z_{j,t_j,out}$ on the source side, the value of the cut will be infinite). In the latter case the value of the cut is increased by $\alpha$, whereas in the former scenario, the value of the cut is increased by {\it at least} $(d - \max_{i\in[s+1]\backslash \{e_j\}} c(i,j))\beta$. This is because any newcomer must be repaired by a cluster differently from his own. As a result, the value of this cut must satisfy
\begin{eqnarray*}
F(Z_{[k],t_0})\ge \sum_{j = 1}^k \min\{(d - \max_{i\in[s+1]\backslash \{e_j\}} c(i,j))\beta,\alpha\}.
\end{eqnarray*}
As discussed in the previous section, in a slightly more general model, one can assume that at any given time-slot a certain number of clusters are unavailable and cannot participate in the repair process. It is not hard to see that this restriction can only increase the value of $F(Z_{[k],t_0})$ as now the maximum is taken over $[s+1]\backslash \Theta$ where $\Theta$ is the set of unavailable clusters. Therefore, the lower-bound above still holds. Let us represent this lower-bound by
\begin{eqnarray}
F^*(e_{[k]}) \stackrel{\triangle}{=} \sum_{j = 1}^k \min\{(d - \max_{i\in[s+1]\backslash \{e_j\}} c(i,j))\beta,\alpha\}.
\label{eqn:bestcut}
\end{eqnarray}
Note that for fixed parameters $\alpha,d$ and $\beta$, the expression in \eqref{eqn:bestcut} is uniquely determined by the sequence $e_{[k]}$, hence the change of the argument in $F^*(\cdot)$ from $Z_{[k],t_0}$ to merely $e_{[k]}$. The first lemma tells us that among all different sequences $e_{[k]}$  the value of $F^*(e_{[k]})$ is minimized when this sequence has a very specific structure.

\begin{lemma}
For any sequence $e_{[k]}\in [s+1]^k$, there exists a sequence $e'_{[k]}\in [2]^k$ such that $F^*(e'_{[k]})\le F^*(e_{[k]})$.
\end{lemma}

\begin{IEEEproof}

 Let $c(i,j)$ be as defined in \eqref{eqn:commuli} and 

\begin{eqnarray*}
e'_j \stackrel{\triangle}{=} \begin{cases}1 &\mbox{ if }  c(e_j,j-1)=\max_{i\in[s+1]} c(i,j-1) \\ 2 &\mbox{ Otherwise. }\end{cases}
\end{eqnarray*}

for $j \in [2:k]$ and $e'_1  \stackrel{\triangle}{=} 1$. Define two variables as follows.

\begin{eqnarray}
v_2(j) \stackrel{\triangle}{=} \begin{cases}0 & \mbox{ if }�j = 0\\ v_2(j-1)+ e'_j-1 &\mbox{ if } 0<j\le k\end{cases}
\label{eqn:cumulative2}
\end{eqnarray}
and 
\begin{eqnarray}
v_1(j) \stackrel{\triangle}{=} j - v_2(j) = \begin{cases}0 & \mbox{ if }�j = 0\\ v_1(j-1)- e'_j+2 &\mbox{ if } 0<j\le k.\end{cases}
\label{eqn:cumulative1}
\end{eqnarray}
For any finite set of integers $D$, let $\max^{(2)}_{i\in D} f(i) = \max_{i\in D\backslash\{x\}} f(i)$ where $x = \argmax_{i\in D}f(i)$. \footnote{if there are multiple maximizers, one can define argmax as the smallest element in $D$ that achieves the maximum.} Also, let $\mathds{1}\{\cdot\}$ be the indicator function. We proceed by proving the following two claims: $v_1(j)=  \max_{i\in[s+1]} c(i,j)$ and $v_2(j) \ge \max^{(2)}_{i\in[s+1]}c(i,j)$. The first claim can be proven by induction. Trivially, $v_1(1) = 1 = \max_i c(i,1) $. Assume the hypothesis is true for $j-1$. Then
\begin{eqnarray*}
v_1(j)&=& v_1(j-1) - e'_j+2 \\
&=&\mathds{1}\{e'_j =1\} (v_1(j-1) + 1) +  \mathds{1}\{e'_j =2\} v_1(j-1) \\
&=& \mathds{1}\{c(e_j,j-1)=\max_{i} c(i,j-1) \} (\max_i c(i,j-1) + 1)\\ &+&  \mathds{1}\{c(e_j,j-1)< \max_{i} c(i,j-1) \} \max_i c(i,j-1)\\
&=&\max_{i} c(i,j).
\end{eqnarray*}
The second claim follows because $v_2(j) = j - v_1(j)  = j - \max_{i} c(i,j) = \sum_{i\neq i^*} c(i,j) \ge \max^{(2)}c(i,j)$ where $i^* = \argmax_i c(i,j)$.

As a result, we have $\max_{i\in[s+1]\backslash\{e_j\}} c(i,j) \le (e'_j-1) v_1(j)+ (2-e'_j)v_2(j)$ for any $j\in[k]$. Therefore,
\begin{eqnarray*}
F^*(e_{[k]})\ge  \sum_{j = 1}^k \min\{(d - (e'_j-1) v_1(j)- (2-e'_j)v_2(j))\beta,\alpha\}.
\end{eqnarray*}

Now consider a sequence of failures and repairs occurring at $t = j\in [k]$ such that if $e'_j = 1$ a server from the first cluster fails and is repaired by the second cluster, and if $e'_j = 2$ then a server from the second cluster fails and is repair by the first cluster. The parameters $v_1(j)$ and $v_2(j)$ represent the number of failed servers up to time-slot $j$, from the first and the second cluster respectively. Therefore we can write
\begin{eqnarray}
F^*(e'_{[k]}) =  \sum_{j = 1}^k \min\{(d - (e'_j-1) v_1(j)- (2-e'_j)v_2(j))\beta,\alpha\},
\label{eqn:binaryfailure}
\end{eqnarray}
thus, $F^*(e'_{[k]}) \le F^*(e_{[k]})$.

\end{IEEEproof}

Suppose now that we are given an arbitrary sequence $e'_{[k]}$ such that $e'_i\in\{1,2\}$. The next question is then how to sort the elements of $e'_{[k]}$ such that the expression in Equation \eqref{eqn:binaryfailure} is minimized. It turns out there is a simple and global answer to this equation. Assume without loss of generality that $v_1(k)\ge v_2(k)$. The next lemma tells us that the failures from each cluster must occur consecutively, without being interrupted. Specifically, $v_1(k)$ servers must fail from the first cluster, and only then, $v_2(k)$ servers fail from the second. Such a pattern always minimizes \eqref{eqn:binaryfailure} regardless of the value of $\alpha$ and $d\beta$.

\begin{lemma}
Let $e'_{[k]}\in[2]^k$ be an arbitrary binary sequence of length $k$ and let $v_1(j)$ and $v_2(j)$ be as defined in equations \eqref{eqn:cumulative1} and \eqref{eqn:cumulative2}. Assume without loss of generality that $v_1(k) \ge v_2(k)$. We have

\begin{eqnarray}
F^*(e'_{[k]}) \ge v_1(k) \alpha+ v_2(k)\min\{(d-v_1(k))\beta,\alpha\}.
\label{eqn:dependsonk1}
\end{eqnarray}
This is achieved with equality if $e'_{[k]}$ is sorted, i.e. if $e'_{[v_1(k)]}$ are all ones.
\end{lemma}
\begin{IEEEproof}
Let us for simplicity define $f(\alpha) = \sum_{j = 1}^k \min\{(d - (e'_j-1) v_1(j)- (2-e'_j)v_2(j))\beta,\alpha\}$ and $g(\alpha) = v_1(k)\alpha+ v_2(k)\min\{(d-v_1(k))\beta,\alpha\}$.
The proof follows from three simple observations.

\begin{itemize}
\item As long as $\alpha \le (d-v_1(k))\beta$ we have  $f(\alpha) = g(\alpha) = k\alpha.$ 
\item The curve $f(\alpha)$ is concave within $ (d-v_1(k))\beta\le \alpha\le d\beta$ whereas the curve $g(\alpha)$ is linear within the same interval. 
\item $f(d\beta) = g(d\beta)$.
\end{itemize}
To see why the last claim holds, note that 
\begin{eqnarray*}
f(d\beta) &=&  \sum_{j = 1}^k (d - (e'_j-1) v_1(j)- (2-e'_j)v_2(j))\beta \\
&=& kd\beta - \beta (\sum_{j= 1}^k  (e'_j-1)  v_1(j)+  (2- e'_j) v_2(j)).
\end{eqnarray*}
Since $g(d\beta) = kd\beta - v_1(k)v_2(k)\beta$, it is left to show that $\sum_{j= 1}^k  (e'_j-1)  v_1(j)+  (2- e'_j) v_2(j) = v_1(k)v_2(k)$. This can be proved by induction over $k$. For $k =1$, the result trivially holds. Let us assume it is true for $k-1$. Then:
\begin{eqnarray*}
\sum_{j = 1}^{k}  (e'_j-1)  v_1(j)+  (2- e'_j) v_2(j) &=& v_1(k-1)v_2(k-1)  + (e'_k-1)  v_1(k)+  (2- e'_k) v_2(k)\\
&=&  (v_1(k)+ e'_k - 2 )(v_2(k)- e'_k  + 1 )  + (e'_k-1)  v_1(k)+  (2- e'_k) v_2(k)\\
&=& v_1(k)v_2(k).
\end{eqnarray*}
\end{IEEEproof}

Remember that $F^*(e'_{[k]})$ is merely a lower bound on the value of the min-cut separating any data collector from the source. But it is easy to find a cut, the value of which is given by Equation \eqref{eqn:dependsonk1}. The sequence of failures and repairs leading to this cut is what is depicted in Figure \ref{fig:twin}: at the end of each time-slot $t\in [0:k_1-1]$ the server $X^{(1)}_{t+1,t}$ fails and is repaired by the second cluster. Next, at the end of each time-slot $t\in[k_1:k-1]$ the server $X^{(2)}_{t-k_1+1,t}$ fails and is repaired by the first cluster. Then a data collector connects to the $k$ servers $X^{(1)}_{[k_1],k}$ and $X^{(2)}_{[k-k_1],k}$. For every $t\in[k_1]$ we include $X^{(1)}_{t,t,in}$ on the source side of the cut and $X^{(1)}_{t,t,out}$ on the sink side. If $(d-k_1)\beta > \alpha$, we do exactly the same thing for the servers in $\{X^{(2)}_{t-k_1,t}| t\in[k_1+1:k]\}$. Otherwise, if $(d-k_1)\beta \le \alpha$ then for all $t\in[k_1+1:k]$ we include both $X^{(2)}_{t-k_1,t,in}$ and $X^{(2)}_{t-k_1,t,out}$ on the sink side.  Since $v_1(k) = k_1$ and $v_2(k) = k - k_1$, the value of this cut is precisely what is given by Equation \eqref{eqn:dependsonk1}.

We have therefore proved the claim which we made at the beginning of this section. The last question to answer is what is the optimal choice of $k_1$ for a specific value of $\alpha$ and $d\beta$. With a slight abuse of notation, let us denote by $F^*(\alpha,d\beta)$ the value of the min-cut for the FCRS with a storage of size $\alpha$ and a repair bandwidth of $\gamma = d\beta$.
\begin{lemma} Suppose we have an FCRS with parameters $n,k,s$ and $d = \lfloor\frac{n}{s}\rfloor$. The value of the min-cut separating any data collector from the source is given by 
\begin{eqnarray}
F^*(\alpha,d\beta) =
\begin{cases}
kd\beta - \lfloor\frac{k}{2}\rfloor\lceil\frac{k}{2}\rceil\beta &\;\;\mbox{ if } \;d\le \frac{\alpha}{\beta},\\
 k_1\alpha + (d-k_1)(k-k_1)\beta&\;\;\mbox{ if } \;d + k - 2k_1 - 1 \le \frac{\alpha}{\beta}<  \min\{d + k - 2k_1 +1,d\} \; \\& \hspace{5cm}\mbox{ for }�k_1 \in[\lceil\frac{k}{2}\rceil : k],\\
 k\alpha &\;\;\mbox{ if } \;\frac{\alpha}{\beta} < d- k - 1.
\end{cases}
\label{eqn:implicit}
\end{eqnarray}
\label{lemma:optimal}
\end{lemma}
\begin{IEEEproof}
Let us define $g(k_1) = k_1\alpha + (k-k_1)\min\{(d-k_1)\beta,\alpha\}$. We want to minimize $g(k_1)$ over $k_1 \in [\lceil\frac{k}{2}\rceil:k ]$ for a specific choice of $\alpha$ and $d\beta$. Without loss of generality we can assume $ (d-k-1)\beta \le \alpha\le d\beta$. If $\alpha >d\beta$ then $g(k_1)$  is clearly minimized  for $k_1 = \lceil \frac{k}{2}\rceil$ which matches with the first line of Equation \eqref{eqn:implicit}. On the other hand, if $\alpha < (d-k-1)\beta $ then $g(k_1)$ is minimized at $k_1 =k$ which yields the last line in \eqref{eqn:implicit}. Let us now minimize a simpler function $h(k_1) = k_1\alpha + (d-k_1)(k-k_1)\beta$. This is a second degree polynomial in $k_1$ and evidently its minimizer over $k_1 \in [\lceil\frac{k}{2}\rceil:k ]$ is $k_1^* = \min\{\lfloor \frac{1}{2}({d+k-\alpha/\beta}) \rceil, k \}$ where $\lfloor\cdot\rceil$ returns the closest integer to its argument. Note that $g(k_1) = \min\{k_1\alpha + (k-k_1)(d-k_1)\beta, k_1\alpha + (k-k_1)\alpha\} =\min\{h(k_1), k\alpha\}  $, so the same $k_1^*$ minimizes $g(\cdot)$ too. Finally, $k_1^* = \min\{\lfloor \frac{1}{2}({d+k-\alpha/\beta}) \rceil, k \}$ implies $d + k - 2k_1^* - 1 \le \frac{\alpha}{\beta}\le d + k - 2k_1^* +1$ which is the same as the second line in Equation \eqref{eqn:implicit}.
\end{IEEEproof}

For any FCRS with parameters $n,k,s$ and file size $M$ we must have $M\le F^*(\alpha,d\beta)$ given by Equation \eqref{eqn:implicit}, otherwise there exists a sequence of failures and repairs after which a data collector (connecting to the newcomers) is incapable of recovering the file. Furthermore, satisfying $M\le  F^*(\alpha,d\beta)$ is sufficient for successfully repairing any sequence of failures, and for any data collector to recover the file, if we resort to random linear codes \cite{ho2006random}. The function $F^*(\alpha,d\beta)$ can be inverted in order to find the minimum value of $\alpha$ for a specific choice of $d\beta$ and $M$, in much the same way as illustrated in \cite{dimakis2010network}. We will sketch this - mostly replicated - proof for the sake of completeness. Let us summarize the result in the next theorem.
\begin{theorem}
The trade-off between $\alpha$ and $\gamma = d\beta$ in an FCRS with parameters $n,k,s$ and $d = \lfloor\frac{n}{s}\rfloor$ can be characterized as
\begin{eqnarray}
\alpha^* &=& \begin{cases} \frac{{M}}{k}& \mbox{ if } \gamma \in [f(0),\infty)\\
\frac{ M - \frac{i}{d}(d- k + i)\gamma}{k-i} & \mbox{ if } \gamma \in [f(i),f(i-1))
\end{cases}
\label{eqn:optimalalpha}
\end{eqnarray}
where
\begin{eqnarray*}
f(i) \stackrel{\triangle}{=} \begin{cases}  \frac{{M}d}{(2k - i - 1)i  +k(d-k+ 1)} \; \; &\mbox{ if }�0\le i< \lfloor\frac{k}{2}\rfloor \\
 \frac{{M}d}{kd - \lfloor\frac{k}{2}\rfloor \lceil\frac{k}{2}\rceil} \; \; &\mbox{ if }�i = \lfloor\frac{k}{2}\rfloor.
\end{cases}
\end{eqnarray*}

\end{theorem}
\begin{IEEEproof}
The function $M = F^*(\alpha,d\beta)$ can be inverted in terms of $\alpha$. 
\begin{eqnarray*}
\alpha^* = F^{-1}(m,d\beta) = \begin{cases}
\frac{M}{k} & \mbox{ if } 0\le M < k(d-k-1)\beta\\
\frac{M - (d-k_1)(k-k_1)\beta}{k_1} & \mbox{ if }  (dk - k_1^2-k_1)\beta 
 \le M < (dk - k_1^2+k_1)\beta \mbox{ for }�k_1 \in[\lceil\frac{k}{2} \rceil + 1: k] \\
\frac{M - (d-\lceil\frac{k}{2}\rceil)\lfloor\frac{k}{2}\rfloor\beta}{\lceil\frac{k}{2}\rceil} & \mbox{ if } (dk - \lceil\frac{k}{2}\rceil^2 -\lceil\frac{k}{2}\rceil ) \beta
 \le M \le (dk - \lfloor\frac{k}{2}\rfloor \lceil\frac{k}{2}\rceil) \beta 
\end{cases}
\end{eqnarray*}
If we write the conditions in terms of $d\beta$ we find
\begin{eqnarray*}
\alpha^* = \begin{cases}
\frac{M}{k} &\mbox{ if } d\beta \ge \frac{dM}{k(d-k-1)}\\
\frac{M - (d-k_1)(k-k_1)\beta}{k_1} &\mbox{ if } d\beta \in [\frac{dM}{dk - k_1^2 + k_1}, \frac{dM}{dk-k_1^2-k_1}] \mbox{ for } k_1 \in[\lceil\frac{k}{2} \rceil + 1: k)\\
\frac{M- (d-\lceil\frac{k}{2}\rceil)\lfloor\frac{k}{2}\rfloor\beta}{\lceil\frac{k}{2}\rceil} & \mbox{ if } d\beta \in [\frac{dM}{dk - \lfloor\frac{k}{2}\rfloor \lceil\frac{k}{2}\rceil}, \frac{dM}{dk - \lceil\frac{k}{2}\rceil^2 - \lceil\frac{k}{2}\rceil} )
\end{cases}
\end{eqnarray*}
which is essentially the same as Equation \eqref{eqn:optimalalpha}. We intentionally substitute $i = k - k_1$ to find an expression similar to Equation (1) in \cite{dimakis2010network}.
\end{IEEEproof}
Let us denote by $(\alpha_{MBR,c},\gamma_{MBR,c})$ the operating point at which the repair bandwidth of the FCRS is minimized. At this point we have 
\begin{eqnarray}
\gamma_{MBR,c} = f(\lfloor\frac{k}{2}\rfloor)= \frac{{M}d}{kd - \lfloor\frac{k}{2}\rfloor \lceil\frac{k}{2}\rceil}.
\label{eqn:mbrfunctional}
\end{eqnarray}
By plugging in this value in Equation \eqref{eqn:optimalalpha} we find
 
\begin{eqnarray*}
\alpha_{MBR,c} = \frac{{M}d}{kd - \lfloor\frac{k}{2}\rfloor \lceil\frac{k}{2}\rceil} = \gamma_{MBR,c}.
\end{eqnarray*}
Therefore,
\begin{eqnarray*}
(\alpha_{MBR,c}, \gamma_{MBR,c}) = ( \frac{d{M}}{kd - \lfloor\frac{k}{2}\rfloor\lceil\frac{k}{2}\rceil}, \frac{d{M}}{kd - \lfloor\frac{k}{2}\rfloor\lceil\frac{k}{2}\rceil}).
\end{eqnarray*}

\subsection{Comparison with  \cite{dimakis2010network}}
\label{sec:functionalcomparison}
\begin{figure}
\centering
\includegraphics[scale= 0.22]{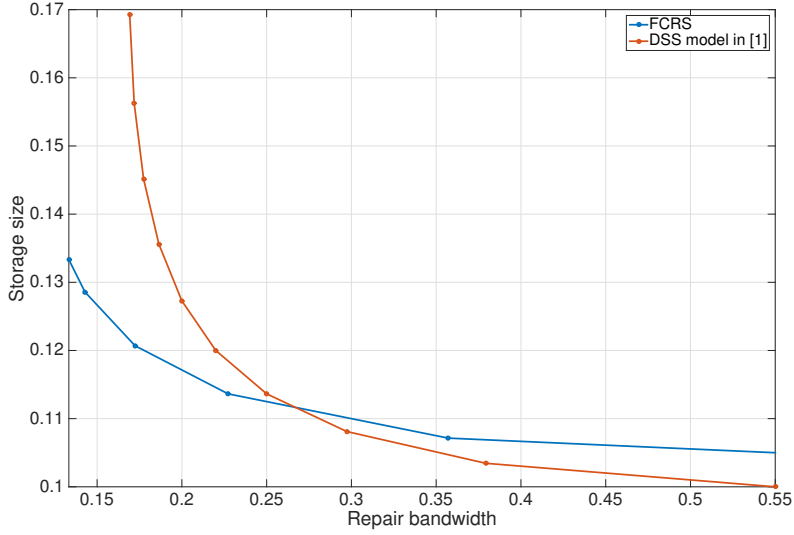}
\caption{Comparison of the achievable region $(\alpha,\gamma)$ for FCRS and \cite{dimakis2010network}. We have $n = 100$, $k = 10$ and both schemes are required to achieve an availability of $s-1 = 9$.}
\label{fig:comparisonfunctional}
\end{figure}

As discussed in the introduction, random linear codes for FCRS can be viewed as an achievability scheme for a more general problem, ACRS, where we are given a DSS and we are required to characterize the region $(\alpha,\gamma)$ for any fixed availability, where $\alpha$ is the storage size and $\gamma$ is the repair bandwidth as defined in Equation \eqref{eqn:gammaacrs}. Interestingly, although the parameter availability has not been a motivation behind the work in \cite{dimakis2010network}, their scheme serves as an achievability result for this general problem too, for any availability $s-1 \in [\lfloor\frac{n-1}{k}\rfloor]$. In particular, the random linear codes proposed in \cite{dimakis2010network} can achieve an availability of $s-1$ if we set $d = d_o = \lfloor\frac{n-1}{s-1}\rfloor$ whereas the random linear codes for FCRS with $s$ complete and one incomplete clusters ($n = ds + s_0$) achieve an availability of $s-1$ with $d = d_c = \lfloor \frac{n}{s}\rfloor $. In this section, we want to illustrate how FCRS can improve the trade-off $(\alpha,\gamma)$ compared to \cite{dimakis2010network} for the same availability and for certain range of parameters. To begin with, we find this comparison most interesting if neither system has any ``residual servers", namely if $s-1| n-1�$ and $s| n$. For instance let us select $n = k^2$ and $s = k$. For FCRS we will have $k$ clusters each of size $k$ and therefore $d_c = k$. For \cite{dimakis2010network} �we have $d_o = \frac{n-1}{s-1} = k + 1$. By plugging in these values of $d$ in Equations \eqref{eqn:optimalalpha} and Equation (1) from \cite{dimakis2010network} respectively, we can find the smallest value of $\alpha$ for any repair bandwidth $\gamma$. This is precisely what we have plotted in Figure \ref{fig:comparisonfunctional} for a choice of $n = 100$ and $k = 10$ (both repair bandwidth and storage size are normalized by $M$). The figure suggests that at small values of repair bandwidth FCRS has a superior performance, and that there is a threshold value of $\gamma$ beyond which it is outperformed by \cite{dimakis2010network}. The improvements offered by FCRS are most visible at the MBR point for which we are going to provide an analytical comparison. Let us define $\gamma_{MBR,o}\stackrel{\triangle}{=}\frac{2d_o{M}}{2kd_o - k^2 + k} $ which is the value of the repair bandwidth at MBR point in \cite{dimakis2010network}.
The two conditions $s|n$ and $s-1|n-1$ imply that $n = (\ell s - \ell + 1)s$ for some positive integer $\ell$. Under this constraint we have $d_o = \ell s + 1$ and $d_c = \ell s  -\ell + 1$ and we can write
\begin{eqnarray*}
\frac{\gamma_{MBR,c}}{\gamma_{MBR,o}} & = &                    \frac{\frac{d_cM}{kd_c - \lfloor\frac{k}{2}\rfloor\lceil\frac{k}{2}\rceil} }{\frac{2d_o{M}}{2kd_o - k^2 + k}}     \le \frac{\frac{d_c{M}}{kd_c - \frac{k^2}{4}}}{\frac{2d_o{M}}{2kd_o - k^2 + k}}  = \frac{\ell s + 1 - (k-1)/2}{\ell s -\ell +  1 - k/4}\cdot \frac{\ell s -\ell + 1}{\ell s +1}.
\end{eqnarray*}
This ratio is upper-bounded by $1$ for almost the entire range of parameters (except when $s = 2$ or $k =1$ or when $(s,k) \in\{(3,2),(3,3),(4,2)\}$) as can be easily verified. The ratio is smallest when $s$ is maximal, that is, $s= \lfloor\frac{n}{k}\rfloor$. This implies $k= \ell s -\ell+ 1 $, which results in

\begin{eqnarray*}
\frac{\gamma_{MBR,c}}{\gamma_{MBR,o}}  \le \frac{2}{3}\cdot \frac{\ell s + \ell + 2}{\ell s +1}.\end{eqnarray*}
This can be upper-bound by 
\begin{eqnarray*}
\frac{\gamma_{MBR,c}}{\gamma_{MBR,o}}\le \frac{2}{3}\cdot \frac{s + 3}{s + 1}
\end{eqnarray*}
which is achieved when $\ell = 1$. This indicates an asymptotic multiplicative improvement of $2/3$ over the repair bandwidth at MBR point in comparison to \cite{dimakis2010network}. \\
It is worth noting that the assumption of ``no residual server"  becomes irrelevant as the parameters $k$ and $s$ grow large and as long as we choose $s = \lfloor\frac{n}{k}\rfloor$.
\begin{proposition}
Let $s$ , $k$ and $s_0$ be three positive integers such that $0\le s_0<\min\{k,s\}$. Let $n = sk +s_0$, $d_c = \lfloor\frac{n}{s}\rfloor = k$ and $d_o = \lfloor\frac{n-1}{s-1}\rfloor$, and define 
\begin{eqnarray*}
f(s,k,s_0) \stackrel{\triangle}{=}\frac{\gamma_{MBR,c}}{\gamma_{MBR,o}} =  \frac{\frac{d_c}{kd_c - \lfloor\frac{k}{2}\rfloor\lceil\frac{k}{2}\rceil}}{\frac{2d_o}{2kd_o - k^2 + k}}.
\end{eqnarray*}
We have
\begin{eqnarray*}
f(s,k,s_0)\le \frac{2}{3}\cdot \frac{(s+3)k + s -3}{(s+1)k -1}.
\end{eqnarray*}
\end{proposition}
\begin{IEEEproof}
\begin{eqnarray*}
f(s,k,s_0) &\le& \frac{\frac{d_c}{kd_c -\frac{k^2}{4}}}{\frac{2d_o}{2kd_o - k^2 + k}}= \frac{d_c}{d_c - \frac{k}{4}} \cdot \frac{2d_o - k + 1}{2d_o}\le \frac{2}{3}\cdot \frac{2d_o - k + 1}{d_o}\\
&\le& \frac{2}{3}(2 - \frac{(k-1)(s-1)}{sk + s_0 -1})\le  \frac{2}{3}(2 - \frac{(k-1)(s-1)}{(s+1)k  -1}) = \frac{2}{3}\cdot \frac{(s+3)k+s-3}{(s+1)k - 1}.
\end{eqnarray*}
\end{IEEEproof}
As a result of this proposition we see that $\lim_{k\rightarrow \infty} f(s,k,s_0) \le (\frac{2}{3}+\epsilon)\cdot \frac{s+3}{s+1}$. If we further let $s\rightarrow \infty$, the ratio of $\frac{2}{3}$ will be established.

Note that this improvement is only achieved asymptotically as $s\rightarrow \infty$ which is mainly of theoretical interest. Nevertheless, FCRS improve the repair bandwidth even when the number of clusters is as small as 3. For instance, if we choose $(n,k,s) = (45,15,3)$, we obtain $\frac{\gamma_{MBR,c}}{\gamma_{MBR,o}}\approx 0.908$.   

\begin{rem}
There is a more intuitive but slightly heuristic approach to compare the performance of the two models. One can  upper-bound the repair bandwidth of the model in \cite{dimakis2010network} as follows
\begin{eqnarray}
\gamma_{MBR,o} = \frac{M}{k}\cdot\frac{1}{1 - \frac{k-1}{2d_o}} < \frac{2M}{k}.
\label{eqn:intuitiveboundo}
\end{eqnarray}
The inequality is due to the fact that $k \le d_o$. As the paramter $s$ grows large, we can get very close to this upper-bound, since $d_o\approx k$. On the other hand, for FCRS we can write
\begin{eqnarray}
\gamma_{MBR,c} =  \frac{M}{k}\cdot\frac{1}{1 - \frac{ \lfloor\frac{k}{2}\rfloor\lceil\frac{k}{2}\rceil}{kd_c}} < \frac{4M}{3k}
\label{eqn:intuitiveboundc}
\end{eqnarray}
which again serves as a good approximation when the number of clusters is large (and consequently $d_c \approx  k$). 
The ratio of these two expressions gives us the same factor of $2/3$. Furthermore, it is evident that no other construction can achieve $\gamma_{MBR} < \frac{M}{k}$. Therefore, an intriguing open question is what is the value of 

\begin{eqnarray*}
c^* = \lim_{n,k\rightarrow \infty}\inf  \frac{\gamma_{MBR}}{M/k}
\end{eqnarray*} 
where the infimum is over all possible functional-recovery constructions that can achieve an availability of $s -1 \approx \frac{n}{k}$.  
\end{rem}
\section{The exact repair model: achievability results for MBR point}
\label{sec:cubic}
In this section we introduce Cubic Codes as a coding scheme designed to minimize the repair bandwidth for the FCRS with $s+1$ clusters where $2\le s\le \lfloor\frac{n}{k}\rfloor$. Cubic Codes are examples of Fractional Repetition Codes \cite{el2010fractional} based on Affine Resolvable Designs \cite{olmez2016fractional}. They can also be viewed as generalizations of grid codes discussed in \cite{olmez2016fractional}. As we shall prove in the next section via a converse bound, Cubic Codes are optimal, in the sense that they minimize the repair bandwidth of FCRS, for 2 and 3 clusters when there is no residual server. On the other hand, they have a strictly worse performance compared to random linear codes that can achieve the cutset bound analyzed in Section \ref{sec:functional}. This implies an inherent gap between the functional repair and exact repair models at the MBR point for the FCRS with 2 and 3 complete clusters. This is by contrast to the DSS model studied in \cite{dimakis2010network} where the MBR point for functional and exact repair coincide. Despite this, we will show that Cubic Codes still achieve an asymptotic multiplicative improvement of $0.79$ on the repair bandwidth compared to the MBR codes \cite{rashmi2009explicit} that guarantee the same availability. We will also prove that if we further restrict ourselves to the repair-by-transfer model, Cubic Codes are optimal for the FCRS with arbitrary parameters.

Suppose the network consists of $n = ds + s_0$ servers divided into $s$ clusters of size $d$ and one cluster of size $s_0 < s$. In this section we further assume that $s_0 < d$. If this is  not true, we can increase $s$ to $s'$ such that $n = ds' + s'_0$ where $s'_0 < \min\{d,s'\}$ and $s'\le \lfloor\frac{n}{k}\rfloor$. Also note that this condition is automatically satisfied if $k^2\ge n$. Assuming the file is large enough, we break it into $m$ independent chunks ${\cal M} = \{{\cal M}_1,\dots,{\cal M}_m\}$ so that $H({\cal M}_i) = M/m$. The value of $m$ will be determined shortly. We start by constructing a $(d^{s+1},m)$ MDS code over these $m$ symbols and indexing the codeword symbols by strings of $s+1$ digits. Let $C_b$ represent a codeword symbol of the MDS code where $b$ is a string of $s+1$ digits, $b = b_{s+1}\dots b_1$, and $b_i\in[d]$.

Server $j$ in Cluster $i$ stores all the codeword symbols of the form $C_b$ where $b_i = j$ and the other indices vary. That is,

\begin{eqnarray*}
X^{(i)}_j  = \{C_b |b_i = j\} \;\;\mbox{ for all } (i,j)\in \{[s]\times[d]\}\cup \{\{s+1\}\times[s_0]\}.
\end{eqnarray*}
\begin{figure}[h]
\centering
\includegraphics[scale=0.8]{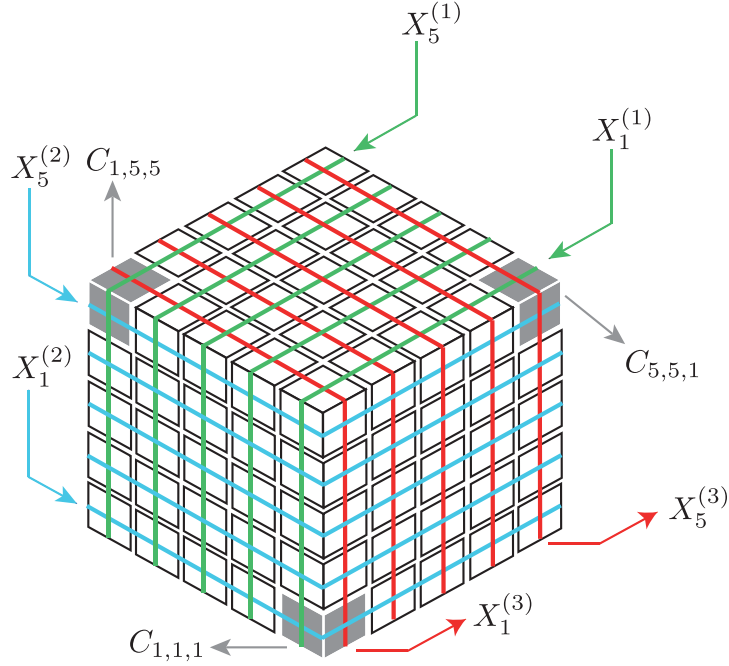}
\caption{Cubic Codes for the FCRS with 3 clusters and with $n = 15$, $d = 5$. The codeword symbols of an MDS code (the small blocks) are arranged in a three dimensional cube. The $j$'th node in the $i$'th cluster stores the codeword symbols that form that $j$'th plane orthogonal to the $i$'th axis.}
\label{fig:cubic}
\end{figure}
This is akin to arranging the codeword symbols of an MDS code in an $s+1$-dimensional hyper-cube and requiring the servers within the $i$'th cluster to store hyperplanes orthogonal to the $i$'th axis. See Figure \ref{fig:cubic} for an illustration. One can also express this code in terms of its generator matrix. Let $B$ be the generator matrix of any $(d^{s+1},m)$ MDS code. For any integer $\ell$ let $\phi_{s+1}(\ell)\dots \phi_1(\ell)$ be the $s+1$-digit expansion of $\ell$ in base $d$, where $\phi_{s+1}(\cdot)$ represents the most significant digit. For any $(i,j)\in \{[s]\times[d]\}\cup \{\{s+1\}\times[s_0]\}$ let $Q^{(i,j)}$ be the $d^s$ by $d^{s+1}$ matrix where 
\begin{eqnarray*}
Q^{(i,j)}_{\ell,r} = \begin{cases}
1 & \mbox{ if } \begin{cases} \phi_e(r) =\phi_e(\ell) & \mbox{ for }e\in[i-1]\mbox{ and } \\ \phi_e(r) = \phi_{e-1}(\ell)  &\mbox{ for } e\in[i+1:s+1] \mbox{ and }\\ \phi_{i}(r) = j-1&\mbox{}\end{cases}\\ 
0 & \mbox{ otherwise. } 
\end{cases}
\end{eqnarray*}
 Then we can write
\begin{eqnarray*}
X^{(i)}_{j} = Q^{(i,j)}B[{\cal M}_1,\dots,{\cal M}_m]^T.
\end{eqnarray*}
If a server $X^{(r)}_\ell$ fails and chooses cluster $i$ for repair, then the $j$'th server in cluster $i$ transmits $Y^{(r,i)}_{\ell,j} = \{C_{b}| b_i = j, b_r = \ell\}$ to the newcomer. Upon receiving all such codeword symbols $Y^{(r,i)}_{\ell,[d]}$, the newcomer is capable of reconstructing the failed server. Furthermore, the newcomer receives a total of $d^{s}$ codeword symbols which shows that for Cubic Codes $d\beta = \alpha$. 

Let us now analyze the performance of this code. Based on the data recovery requirement, we know that every $k$ servers in the network, regardless of their cluster must be able to recover the file. Consider a set of $k$ servers chosen in such a way that $k_i$ servers belong to cluster $i$ where $\sum_{i=1}^{s+1}{k_i} = k$. These servers together provide a total of $R$ codeword symbols of the MDS code where
\begin{eqnarray*}
R &=& \sum_{i=1}^{s+1} d^{s}k_i - \sum_{i=1}^{s+1}\sum_{j= i+1}^{s+1} d^{s-1}k_i k_j\\ &+& \sum_{i=1}^{s+1}\sum_{j= i+1}^{s+1}\sum_{\ell = j+1}^{s+1} d^{s-2}k_i k_j k_\ell -\dots\\
&=& d^{s+1} - \prod_{i=1}^{s+1}(d-k_i).
\end{eqnarray*}
Thus, in order for the file ${\cal M}$ to be recoverable from these $k$ servers, we must have
\begin{eqnarray}
d^{s+1} - \prod_{i=1}^{s+1}(d-k_i) \ge m.
\label{eqn:boundedm}
\end{eqnarray}
Note that this inequality must be true for any choice of the parameters $k_i$. Let us therefore minimize the left hand side of this inequality over the constraints $\sum_{i=1}^{s+1}k_i = k$, $k_i \ge 0$ and $k_{s+1}\le s_0$.

\begin{eqnarray}
k^*_{[s+1]}  = \argmin_{k_{[s+1]}:\sum_{i=1}^{s+1}k_i = k\mbox{   and   }k_{s+1}\le s_0} d^{s+1} - \prod_{i=1}^{s+1}(d-k_i).
\label{eqn:bestk}
\end{eqnarray}

To solve this optimization problem, it is necessary to distinguish between two regimes. 
\begin{itemize}
\item {\bf Regime 1:} $s_0 \ge  \lfloor{\frac{k}{s +1}}\rfloor$.

 Define $s_1 = k \mod s+1$. The solution to \eqref{eqn:bestk} can be expressed as
 \begin{eqnarray*}
k^*_{i}  = \begin{cases} 
\lceil{\frac{k}{s+1}}\rceil & \mbox{ if }   i\le s_1\\
\lfloor{\frac{k}{s+1}}\rfloor &  \mbox{ if }  s_1 < i\le s + 1.
\end{cases}
\end{eqnarray*}
Returning to Inequality \eqref{eqn:boundedm} we can write $m =  d^{s+1} -  (d - \lceil{\frac{k}{s+1}}\rceil)^{s_1} (d - \lfloor{\frac{k}{s+1}}\rfloor)^{(s +1- s_1)}$. Since each server stores $d^{s}$ codeword symbols, the storage size is 

\begin{eqnarray*}
\alpha = \frac{Md^s}{m} = \frac{M�d^{s}}{ d^{s+1} -  (d - \lceil{\frac{k}{s+1}}\rceil)^{s_1} (d - \lfloor{\frac{k}{s+1}}\rfloor)^{(s+1- s_1)}}.
\end{eqnarray*}

We saw that for Cubic Codes $\gamma = \alpha$. Therefore,
\begin{eqnarray*}
\gamma  =\frac{M�d^{s}}{ d^{s+1} -  (d - \lceil{\frac{k}{s+1}}\rceil)^{s_1} (d - \lfloor{\frac{k}{s+1}}\rfloor)^{(s +1- s_1)}}.
\end{eqnarray*}
\item {\bf Regime 2:} $s_0 <  \lfloor{\frac{k}{s +1}}\rfloor$.

This time define $s_1 = k-s_0\mod s$. The solution to \eqref{eqn:bestk} is 
\begin{eqnarray*}
k^*_{i}  = \begin{cases} 
\lceil{\frac{k-s_0}{s}}\rceil & \mbox{ if } i\le s_1\\
\lfloor{\frac{k-s_0}{s}}\rfloor &  \mbox{ if } s_1 < i\le s \\
s_0&  \mbox{ if }   i = s + 1.
\end{cases}
\end{eqnarray*}
Again, plugging this into \eqref{eqn:boundedm} we have
\begin{eqnarray*}
m &=& d^{s+1} -  (d - \lceil{\frac{k-s_0}{s}}\rceil)^{s_1}(d - \lfloor{\frac{k-s_0}{s}}\rfloor)^{s+1-s_1}\\&-& ( \lfloor{\frac{k-s_0}{s}}\rfloor - s_0)(d -\lceil{\frac{k-s_0}{s}}\rceil)^{s_1}(d -\lfloor{\frac{k-s_0}{s}}\rfloor)^{s-s_1}\\
&=&d^{s+1} -  (d-s_0)(d - \lceil{\frac{k-s_0}{s}}\rceil)^{s_1}(d - \lfloor{\frac{k-s_0}{s}}\rfloor)^{s-s_1}.
\end{eqnarray*}

Consequently,

\begin{eqnarray*}
\gamma = \alpha = \frac{{M}d^s}{d^{s+1} -  (d-s_0)(d - \lceil{\frac{k-s_0}{s}}\rceil)^{s_1}(d - \lfloor{\frac{k-s_0}{s}}\rfloor)^{s-s_1}}.
\end{eqnarray*}

Let us summarize this in the following thereom.
\end{itemize}
\begin{theorem}
Suppose we have an FCRS with parameters $n,k,s$ where $n = ds + s_0$ and $s_0 < \min\{d,s\}$. The Cubic Codes achieve a repair bandwidth of

\begin{eqnarray}
\gamma_{cc}(n,k,s) =
\begin{cases}
\frac{{M}�d^{s}}{ d^{s+1} -  (d - \lceil{\frac{k}{s+1}}\rceil)^{s_1} (d - \lfloor{\frac{k}{s+1}}\rfloor)^{(s +1- s_1)}} & \mbox{ if } s_0 \ge  \lfloor{\frac{k}{s +1}}\rfloor\\
\frac{{M}d^s}{d^{s+1} -  (d-s_0)(d - \lceil{\frac{k-s_0}{s}}\rceil)^{s_2}(d - \lfloor{\frac{k-s_0}{s}}\rfloor)^{s-s_2}} & \mbox{ if } s_0 <  \lfloor{\frac{k}{s +1}}\rfloor
\end{cases}
\label{eqn:gammacc}
\end{eqnarray}
where $s_1 = k \mod s+1$, and $s_2 = k-s_0 \mod s$.
\end{theorem} 
An interesting regime is when there are no residual servers, that is when $n = sd$. In this case we have
\begin{eqnarray}
\gamma_{cc}(sd,k,s) =
\frac{{M}/d}{1-(1 - \lceil{\frac{k}{s}}\rceil/d)^{s_2}(1 - \lfloor{\frac{k}{s}}\rfloor/d)^{s-s_2}}.
\label{eqn:comelater}
\end{eqnarray}
This can be further simplified if we assume $s|k$.
\begin{eqnarray*}
\gamma_{cc}(sd,\ell s, s) =
\frac{{M}/d}{1-(1 -\frac{\ell}{d})^{s}}.
 \end{eqnarray*}
 In fact, it follows from a simple argument that
 \begin{eqnarray}
\gamma_{cc}(sd,k,s) \le \frac{{M}/d}{1-(1 -\frac{k}{sd})^{s}}.
\label{eqn:someapprox}
 \end{eqnarray}
 To see why, note that 
  \begin{eqnarray*}
 \left( \frac{1- \frac{\lceil k/s\rceil}{d}}{1- \frac{k}{sd}}\right)^{s_2} \left( \frac{1- \frac{\lfloor k/s\rfloor}{d}}{1- \frac{k}{sd}}\right)^{s - s_2}\le 1
  \end{eqnarray*} 
which is true since the geometric mean of $s$ numbers is upperbounded by their arithmetic mean.\\

It is not difficult to see that if we fix $s$ and $k$, $\gamma_{cc}(n,k,s)$ is monotonically decreasing in $n$. This is because adding one more server to the last cluster cannot increase the expression $\min_{k_{[s+1]}} d^{s+1}-\prod_{i=1}^{s+1} (d-k_i)$. Based on this property and Equation \eqref{eqn:someapprox} we can establish the following bound. 
\begin{corollary} Let $\gamma_{cc}(n,k,s)$ be the repair bandwidth of Cubic Codes for an FCRS with parameters $n,k,s$  where $n = ds + s_0$ and $s_0\le\min\{s,d\}$. Then
 \begin{eqnarray*} 
 \gamma_{cc}(n,k,s) \le \frac{{M}/d}{1-(1 -\frac{k}{(s+1)d})^{s+1}}.
  \end{eqnarray*}
  \end{corollary}
This bound is sufficiently tight for our purpose and we will resort to it for our analytical comparison in the next section.

\subsection{Comparison with Functional Repair and  \cite{rashmi2009explicit}}
\label{sec:cubiccomparison}
\begin{figure}[h]
\centering
\includegraphics[scale=0.2]{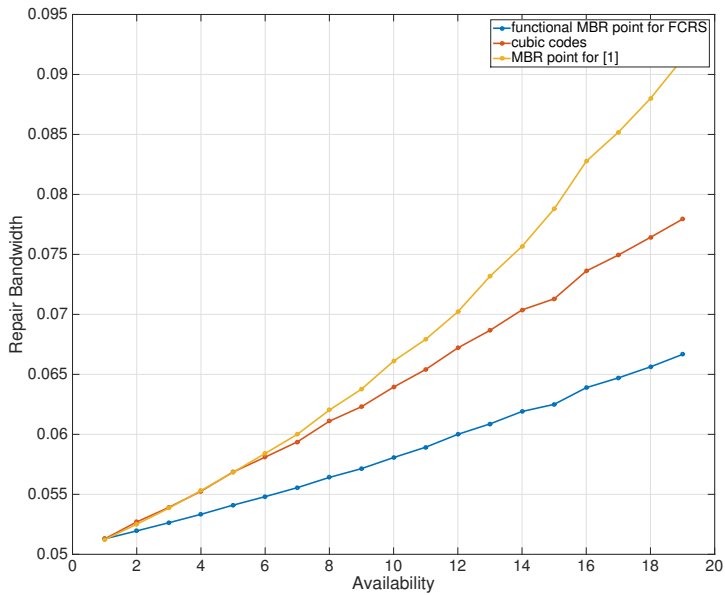}
\caption{Comparison of Cubic Codes with the functional repair MBR point  of FCRS and the MBR point of \cite{dimakis2010network}.}
\label{fig:compare_mbr}
\end{figure}

Let us start with a numerical comparison. Here we fix the number of servers $n$ and let the availability grow gradually. For every fixed availability we compare the repair bandwidth for the three schemes: the MBR point for functional repair of FCRS in Section \ref{sec:functional},  that is expression \eqref{eqn:mbrfunctional}, the MBR codes proposed in \cite{rashmi2009explicit} (which corresponds to the functional repair MBR point in \cite{dimakis2010network}) and finally the Cubic Codes, that is expression \eqref{eqn:gammacc}.  For this numerical analysis we set $n = 400$,  $k = 20$, and we let $s-1$ grow from $1$ to $19$. We normalize the repair bandwidth by the size of the file. As can be seen in Figure \ref{fig:compare_mbr}, as $s$ grows large Cubic Codes perform somewhere in between the functional repair points of FCRS and \cite{dimakis2010network}.  The multiplicative improvement over \cite{dimakis2010network}� can be measured around $0.79$ at its peak, i.e. when $s -1 = 19$. This will be theoretically justified next.

Let us first bound the ratio of repair bandwidth for Cubic Codes and functional repair bandwidth for the FCRS. Assuming $k$ even we have

\begin{eqnarray*}
\frac{\gamma_{cc}}{\gamma_{MBR,c}} & \le &\frac{\frac{{M}/d_c}{1-(1 -\frac{k}{(s+1)d_c})^{s+1}}
}{\frac{{M}}{k - \frac{k^2}{4d_c}}}  = \frac{k(1 - \frac{k}{4d_c})}{d_c - d_c(1- \frac{k}{(s+1)d_c})^{s+1}}\\&\le& \frac{k}{d_c}(1-\frac{k}{4d_c})\frac{1}{1-e^{-k/d_c}}\le\frac{3}{4(1-e^{-1})} .
 \end{eqnarray*}
In conjunction with the results of Section \ref{sec:functionalcomparison},  if we further assume that $s$ is chosen as large as possible, i.e. $s = \lfloor\frac{n}{k}\rfloor$, we can write

  \begin{eqnarray*}
\frac{\gamma_{cc}}{\gamma_{MBR,o}}\le \frac{3}{4(1-e^{-1})} \cdot   \frac{2}{3}\cdot \frac{(s+3)k + s -3}{(s+1)k -1}.
 \end{eqnarray*}
Since the expression for the ratio $\frac{\gamma_{cc}}{\gamma_{MBR,o}}$ does not depend on whether $k$ is odd or even, the same bound holds for general $k$.  We have therefore established the following proposition.
\begin{proposition}
Let $s$ , $k$ and $s_0$ be three positive integers such that $s_0<\min\{k,s\}$. Let $n = sk + s_0$, $d_c = \lfloor\frac{n}{s}\rfloor = k$ and $d_o = \lfloor\frac{n-1}{s-1}\rfloor$, and define 
\begin{eqnarray*}
g(s,k,s_0)\stackrel{\triangle}{=} \frac{\gamma_{cc}}{\gamma_{MBR,o}} & \le &\frac{\frac{1}{1-(1 -\frac{k}{(s+1)d_c})^{s+1}}
}{\frac{2d_o}{2kd_o - k^2 + k}}.
 \end{eqnarray*}
 We have
\begin{eqnarray*}
g(s,k,s_0)\le \frac{1}{2(1-e^{-1})}\cdot \frac{(s+3)k + s -3}{(s+1)k -1}.
\end{eqnarray*}
\end{proposition}
Based on this proposition, if we let $k\rightarrow\infty$, we find
  \begin{eqnarray*}
\frac{\gamma_{cc}}{\gamma_{MBR,o}}\le \frac{3}{4(1-e^{-1})} \cdot (\frac{2}{3}+\epsilon)\cdot \frac{s +3 }{s + 1} \approx 0.79 \cdot \frac{s +3 }{s + 1}.
  \end{eqnarray*}
This proves that for the same availability of $s -1 = \lfloor\frac{n}{k}\rfloor -1$, Cubic Codes achieve an asymptotic (as $k,s,n \rightarrow \infty$) multiplicative improvement of $0.79$ over the minimum repair bandwidth in comparison to MBR point in \cite{dimakis2010network}. Note that this improvement is only achieved asymptotically as $s\rightarrow \infty$ which is mainly of theoretical interest. Nevertheless, Cubic Codes improve the repair bandwidth even when the number of clusters is as small as 3. For instance, if we choose $(n,k,s) = (45, 15, 3)$, we obtain $\frac{\gamma_{cc}}{\gamma_{MBR,o}}\approx 0.97$.   
\begin{rem}
We can further upper-bound the value of the repair bandwidth of Cubic Codes as follows\begin{eqnarray*}
\gamma_{MBR,cc} \le  \frac{M}{d}\cdot\frac{1}{1 - (1 - \frac{k}{d(s + 1)})^{s + 1}} \le \frac{M}{k}\cdot\frac{1}{1-(1-\frac{1}{s+1})^{s+1}}\le \frac{1.58M}{k}.
\end{eqnarray*}
One could gain intuition by dividing this to the upper-bound in Equation \eqref{eqn:intuitiveboundo} which results in the same factor of $0.79$. Here we ask the same open question as in the last section. What is the value of 

\begin{eqnarray*}
c^* = \lim_{n,k\rightarrow \infty}\inf  \frac{\gamma_{MBR}}{M/k}
\end{eqnarray*} 
where the infimum is over all possible exact-recovery constructions that can achieve an availability of $s -1 \approx \frac{n}{k}$.
\end{rem}

\section{ converse bound for exact repair}
\label{sec:converse}

In this section we provide an  exact repair converse bound for the FCRS. The main purpose of this converse bound is to prove that the Cubic Codes introduced in Section \ref{sec:cubic} minimize the repair bandwidth for the FCRS with two or three complete clusters and no residual servers. As a consequence of this, a fundamental gap between the exact and functional repair regimes at the MBR point is established for FCRS. This is by contrast to the DSS model studied in \cite{dimakis2010network} where the MBR point for functional and exact repair coincide.
Unfortunately a straightforward generalization of the bound to more than three clusters is loose and is omitted for this reason. However, we can prove that Cubic Codes are optimal for the FCRS with arbitrary parameters under the more restricted repair-by-transfer model. It is noteworthy that the only property of  repair-by-transfer that we use in our proof is the fact that $H( Y^{(r,i)}_{\ell,[d]} | X^{(r)}_\ell ) = 0$. It is quite tempting to conjecture that this same property holds more generally, at the MBR point for FCRS under the exact repair model. If this conjecture is true, then Cubic Codes are optimal for FCRS with arbitrary parameters under  exact repair.

\begin{theorem}
Cubic Codes achieve the minimum repair bandwidth under exact repair for the FCRS with $2$ or $3$ complete clusters, i.e. when $s\in \{2,3\}$ and $s_0 = 0$.
\label{thm:ssmall}
\end{theorem}
\begin{theorem}
Cubic Codes achieve the minimum repair bandwidth under the repair-by-transfer model for the FCRS with arbitrary parameters.
\label{thm:rbt}
\end{theorem}
\begin{IEEEproof}[Proof of theorem \ref{thm:ssmall}]
We will present the proof for $s = 3$. The proof for $s = 2$ is omitted to avoid redundancy. Therefore, we have $n = 3d$. Assume $k_i$ servers from the $i$'th cluster take part in the data recovery. Specifically, let $\tau_{i}\subseteq [d]$ for $i\in[3]$ represent the set of indices of the servers from $i$'th cluster that are connected to a data collector, such that $|\tau_i| = k_i$ and $\sum_{i=1}^3 k_i = k$. By taking an average over all possible such choices of $\tau_1,\tau_2,\tau_{3}$ we can write 

\begin{eqnarray}
M&\le& \frac{1}{{d \choose k_1}{d \choose k_2}{d \choose k_3}}\sum_{\stackrel{\tau_1,\tau_2,\tau_3}{\tau_i\subseteq[d],\;|\tau_i|= k_i}}H(X^{(1)}_{\tau_1},X^{(2)}_{\tau_2},X^{(3)}_{\tau_3})\nonumber\\
&=& \frac{1}{{d \choose k_1}{d \choose k_2}{d \choose k_3}}\sum_{{\tau_1,\tau_2,\tau_3}}H(X^{(3)}_{\tau_3}|X^{(2)}_{\tau_2},X^{(1)}_{\tau_1}) + \frac{1}{{d \choose k_1}{d \choose k_2}}\sum_{{\tau_1,\tau_2}}H(X^{(2)}_{\tau_2}|X^{(1)}_{\tau_1})\nonumber \\&+& \frac{1}{{d \choose k_1}}\sum_{{\tau_1}}H(X^{(1)}_{\tau_1}).
\label{eqn:eq11}
\end{eqnarray}
Let us start by upper-bounding the first term.
\begin{eqnarray*}
Q_3 &\stackrel{\triangle}{=}&  \frac{1}{{d \choose k_1}{d \choose k_2}{d \choose k_3}}\sum_{{\tau_1,\tau_2,\tau_3}}H(X^{(3)}_{\tau_3}|X^{(2)}_{\tau_2},X^{(1)}_{\tau_1}) \stackrel{(*)}{\le}  \frac{1}{{d \choose k_1}{d \choose k_2}{d \choose k_3}}\sum_{{\tau_1,\tau_2,\tau_3}}H(Y^{(3,2)}_{\tau_3,[d]}|Y^{(3,2)}_{\tau_3,\tau_2},X^{(1)}_{\tau_1}) \\
 &= &  \frac{1}{{d \choose k_1}{d \choose k_2}{d \choose k_3}}\sum_{{\tau_1,\tau_2,\tau_3}}H(Y^{(3,2)}_{\tau_3,[d]\backslash\tau_2}|Y^{(3,2)}_{\tau_3,\tau_2},X^{(1)}_{\tau_1})
 \stackrel{(\#)}{\le}  \frac{1}{{d \choose k_1}{d \choose k_3}}\sum_{{\tau_1,\tau_3}}\frac{d-k_2}{d}H(Y^{(3,2)}_{\tau_3,[d]}|X^{(1)}_{\tau_1}).
  \end{eqnarray*}
Inequality $(*)$ follows from the fact that $H(X^{(3)}_{\tau_3}|Y^{(3,2)}_{\tau_3,[d]}) =0$ and $H(Y^{(3,2)}_{\tau_3,\tau_2}|X^{(2)}_{\tau_2}) = 0 $. Inequality $(\#)$ follows from (conditional) Han's inequality.
 Let us continue by bounding the right hand side of  this inequality.
\begin{eqnarray*}
Q_3&\le& \frac{1}{{d \choose k_1}{d \choose k_3}}\sum_{{\tau_1,\tau_3}}\frac{d-k_2}{d}H(Y^{(3,2)}_{\tau_3,[d]}|X^{(1)}_{\tau_1})\\ &=&  \frac{1}{{d \choose k_1}{d \choose k_3}}\sum_{{\tau_1,\tau_3}}\frac{d-k_2}{d}H(X^{(3)}_{\tau_3}|X^{(1)}_{\tau_1})+ \frac{1}{{d \choose k_1}{d \choose k_3}}\sum_{{\tau_1,\tau_3}}\frac{d-k_2}{d}H(Y^{(3,2)}_{\tau_3,[d]}|X^{(1)}_{\tau_1},X^{(3)}_{\tau_3})\\
 &\le& \frac{1}{{d \choose k_1}{d \choose k_3}}\sum_{{\tau_1,\tau_3}}\frac{d-k_2}{d}H(Y^{(3,1)}_{\tau_3,[d]}|Y^{(3,1)}_{\tau_3,\tau_1})+ \frac{1}{{d \choose k_1}{d \choose k_3}}\sum_{{\tau_1,\tau_3}}\frac{d-k_2}{d}H(Y^{(3,2)}_{\tau_3,[d]}|X^{(3)}_{\tau_3})\\
 &\stackrel{(*)}{\le}& \frac{1}{{d \choose k_3}} \frac{(d-k_2)(d-k_1)}{d^2}\sum_{\tau_3}H(Y^{(3,1)}_{\tau_3,[d]})+ k_3(d-k_2)\beta - \frac{d-k_2}{d{d\choose k_3}}\sum_{\tau_3}H(X^{(3)}_{\tau_3})\\
 &\le& k_3\frac{(d-k_2)(d-k_1)}{d}\beta + k_3(d-k_2)\beta -\frac{d-k_2}{d{d\choose k_3}}\sum_{\tau_3}H(X^{(3)}_{\tau_3}) .
\end{eqnarray*}
Inequality $(*)$ follows from the fact that $H(Y|X) = H(Y) - H(X)$ if $X$ is a function of $Y$ and from applying Han's inequality for a second time. We go back to Equation \eqref{eqn:eq11} and bound the second term.
\begin{eqnarray*}
Q_2 &\stackrel{\triangle}{=}&\frac{1}{{d \choose k_1}{d \choose k_2}}\sum_{{\tau_1,\tau_2}}H(X^{(2)}_{\tau_2}|X^{(1)}_{\tau_1}) \le \frac{1}{{d \choose k_1}{d \choose k_2}}\sum_{{\tau_1,\tau_2}}H(Y^{(2,1)}_{\tau_2,[d]}|Y^{(2,1)}_{\tau_2,\tau_1})\\
&=&\frac{1}{{d \choose k_1}{d \choose k_2}}\sum_{{\tau_1,\tau_2}}H(Y^{(2,1)}_{\tau_2,[d]\backslash{\tau_1}}|Y^{(2,1)}_{\tau_2,\tau_1})\\
&\le& \frac{1}{{d \choose k_2}}\frac{d-k_1}{d}\sum_{\tau_2}H(Y^{(2,1)}_{\tau_2,[d]})\le k_2(d-k_1)\beta.
\end{eqnarray*}
Therefore, we proved
\begin{eqnarray*}
M&\le& Q_3 + Q_2 + \frac{1}{{d \choose k_1}}\sum_{{\tau_1}}H(X^{(1)}_{\tau_1}) \\
&\le& k_3\frac{(d-k_2)(d-k_1)}{d}\beta +k_3(d-k_2)\beta-\frac{d-k_2}{d{d\choose k_3}}\sum_{\tau_3}H(X^{(3)}_{\tau_3})\\ &+& k_2(d-k_1)\beta + \frac{1}{{d \choose k_1}}\sum_{\tau_1}H(X^{(1)}_{\tau_1}).
\end{eqnarray*}
Via an identical procedure we can more generally prove that if $\pi:\{1,2,3\}\rightarrow\{1,2,3\}$ is a bijection, then
\begin{eqnarray*}
M&\le& k_{\pi_3}\frac{(d-k_{\pi_2})(d-k_{\pi_1})}{d}\beta +k_{\pi_3}(d-k_{\pi_2})\beta-\frac{d-k_{\pi_2}}{d{d\choose k_{\pi_3}}}\sum_{\tau_{\pi_3}}H(X^{(\pi_3)}_{\tau_{\pi_3}}) \\&+& k_{\pi_2}(d-k_{\pi_1})\beta + \frac{1}{{d \choose k_{\pi_1}}}\sum_{\tau_{\pi_1}}H(X^{({\pi_1})}_{\tau_{\pi_1}}).
\end{eqnarray*}
By averaging this inequality over all possible bijections $\pi$ we find
\begin{eqnarray*}
M&\le& k_3\frac{(d-k_2)(d-k_1)}{3d}\beta  + k_2\beta\frac{(d-k_3)(d-k_1)}{3d} + k_1\beta\frac{(d-k_3)(d-k_1)}{3d} \nonumber\\
&+& \frac{1}{3}\left((d-k_1)k_2 +(d-k_1)k_3 +(d-k_2)k_1 +(d-k_2)k_3 +(d-k_3)k_1 +(d-k_3)k_2\right)\beta\nonumber\\
&+&\frac{1}{3}(k_1k_2 + k_1k_3+ k_2k_3)\beta\nonumber\\
&=& \frac{\beta}{3}(2dk +  k_3\frac{(d-k_2)(d-k_1)}{d}  + k_2\frac{(d-k_3)(d-k_1)}{d} + k_1\frac{(d-k_2)(d-k_3)}{d}-k_1k_2-k_1k_3-k_2k_3)\\
&=& \frac{\beta}{3d}(2d^2k  + 3k_1k_2k_3 -3d(k_1k_2 + k_1k_3 + k_2k_3) + d^2k)\\
&=& \frac{\beta}{d}(d^3 - (d-k_1)(d-k_2)(d-k_3)),
\label{eqn:ineq16}
\end{eqnarray*}
where we have used the fact that $\alpha \le {d\beta}$ (due to the repair requirement) and $\sum{k_i} = k$. Note that the inequality above must hold for any choice of $k_1,k_2,k_3$ that satisfy $\sum{k_i} = k$. In particular we must be able to choose $k_i = \lceil\frac{k}{3}\rceil$ for $i\in[s_2]$ and $k_i = \lfloor\frac{k}{3}\rfloor$ for $i\in[s_2+1:3]$ where $s_2 =k\mod 3$. For this choice of $k_{[3]}$ we have

�\begin{eqnarray*}
\gamma = d\beta \ge \frac{Md^2}{d^3 - (d-\lceil\frac{k}{3}\rceil)^{s_2} (d-\lfloor\frac{k}{3}\rfloor)^{3-s_2}} = \frac{M/d}{1 - (1-\lceil\frac{k}{3}\rceil/d)^{s_2} (1-\lfloor\frac{k}{3}\rfloor/d)^{3-s_2}}. 
\end{eqnarray*}
This is the same expression as the achievable repair bandwidth of Cubic Codes specified by Equation \eqref{eqn:comelater} if we set $s = 3$.
\end{IEEEproof}
\begin{IEEEproof}[Proof of theorem \ref{thm:rbt}]
The only property of the repair-by-transfer model which we use is \begin{eqnarray}
H( Y^{(r,i)}_{\ell,[d]} | X^{(r)}_\ell ) = 0.
\label{eqn:theonlyrequirement}
\end{eqnarray}
 Suppose $n = sd + s_0$.  Assume $k_i$ servers from the $i$'th cluster take part in the data recovery. Specifically, let $\tau_{i}\subseteq [d]$ for $i\in[s]$ and $\tau_{s+1}\subseteq [s_0]$  represent the set of indices of the servers from each cluster that are connected to a data collector, such that $|\tau_i| = k_i$ and $\sum_{i=1}^{s+1} k_i = k$. By taking an average over all possible such choices of $\tau_1,\dots,\tau_{s+1}$ we can write 

\begin{eqnarray}
M&\le& \frac{1}{\prod_{i =1}^{s}{d \choose k_i}{s_0 \choose k_{s+1}}}\sum_{\tau_1,\dots,\tau_{s+1}}H(X^{(1)}_{\tau_1},\dots,X^{(s+1)}_{\tau_{s+1}})\nonumber\\
&=& \sum_{j =1}^{s}\frac{1}{\prod_{i =1}^j{d \choose k_i}}\sum_{{\tau_1,\dots,\tau_j}}H(X^{(j)}_{\tau_j}|X^{(j-1)}_{\tau_{j-1}}\dots,X^{(1)}_{\tau_1})\\
&+& \frac{1}{\prod_{i =1}^{s}{d \choose k_i}{s_0\choose k_{s+1}}}\sum_{{\tau_1,\dots,\tau_{s+1}}}H(X^{(s+1)}_{\tau_{s+1}}|X^{(s)}_{\tau_{s}}\dots,X^{(1)}_{\tau_1}).
\label{eqn:sadasdqw}
\end{eqnarray}
Let us define
\begin{eqnarray*}
Q_j  = \frac{1}{\prod_{i =1}^j{d \choose k_i}}\sum_{{\tau_1,\dots,\tau_j}}H(X^{(j)}_{\tau_j}|X^{(j-1)}_{\tau_{j-1}}\dots,X^{(1)}_{\tau_1}) \;\; j\in [s],
\end{eqnarray*}
and 
\begin{eqnarray*}
Q_{s+1} = \frac{1}{\prod_{i =1}^s{d \choose k_i}{s_0\choose k_{s+1}}}\sum_{{\tau_1,\dots,\tau_{s+1}}}H(X^{(s+1)}_{\tau_{s+1}}|X^{(s)}_{\tau_{s}}\dots,X^{(1)}_{\tau_1}).
\end{eqnarray*}
We have
\begin{eqnarray*}
Q_j &\le& \frac{1}{\prod_{i =1}^j{d \choose k_i}}\sum_{{\tau_1,\dots,\tau_j}}H(Y^{(j,j-1)}_{\tau_j,[n]}|Y^{(j,j-1)}_{\tau_j,\tau_{j-1}},X^{(j-2)}_{\tau_{j-2}},\dots,X^{(1)}_{\tau_1})\\
&\le& \frac{1}{\prod_{i =1}^{j-1}{d \choose k_i}}\frac{d-k_{j-1}}{d}H(Y^{(j,j-1)}_{\tau_j,[n]}|X^{(j-2)}_{\tau_{j-2}}\dots,X^{(1)}_{\tau_1})\\
&=& \frac{1}{\prod_{i =1}^{j-1}{d \choose k_i}}\frac{d-k_{j-1}}{d}H(X^{(j)}_{\tau_j}|X^{(j-2)}_{\tau_{j-2}}\dots,X^{(1)}_{\tau_1})\\
&\le&\dots\\
&\le& \left(\prod_{i = 1}^{j-1} \frac{d-k_i}{d}\right) H(X^{(j)}_{\tau_j})\\
&\le& \left(\prod_{i = 1}^{j-1} \frac{d-k_i}{d}\right) k_jd\beta.
\end{eqnarray*}
Similarly, we can establish 
\begin{eqnarray*}
Q_j &\le& \left(\prod_{i = 1}^{s} \frac{d-k_i}{d}\right) k_{s+1}d\beta.
\end{eqnarray*}
Therefore we have
\begin{eqnarray*}
M&\le& \sum_{j = 1}^{s+1} Q_j \le \sum_{j = 1}^{s+1}  \left(\prod_{i = 1}^{j-1} \frac{d-k_i}{d}\right) k_jd\beta= d^2\beta\left(1 - \prod_{i = 1}^{s+1} (1 - \frac{k_i}{d})\right)
\end{eqnarray*}
and as a result
\begin{eqnarray}
\gamma(n,k,s) = d\beta \ge \frac{Md^s}{d^{s+ 1} - \prod_{i = 1}^{s+1}(d-k_i)}.
\label{eqn:againn}
\end{eqnarray}
This must hold true for any choice of $k_{[1:s+1]}$ that satisfies $\sum_{i = 1}^{s+1}{k_i} = k$. In particular we must be able to choose $k_{[1:s+1]} = k^*_{[1:s+1]}$ where $k^*_{[1:s+1]}$ is described in Regimes 1 and 2 in Section \ref{sec:cubic}. Plugging this in \eqref{eqn:againn} we obtain 

\begin{eqnarray}
\gamma(n,k,s)\ge \gamma_{cc}(n,k,s).
\end{eqnarray}
where $\gamma_{cc}(n,k,s)$ is the repair bandwidth of Cubic Codes provided by \eqref{eqn:gammacc}.
\end{IEEEproof}

\section{Conclusion and Open Problems}

\label{sec:conclusion}
In this work we proposed FCRS as a distributed storage architecture which achieves high availability and low repair bandwidth. We demonstrated that in the functional repair and exact repair paradigms and for the same availability, FCRS can improve the repair bandwidth by asymptotic multiplicative factors of $2/3$ and $0.79$ (respectively) compared to the literature. Several intriguing questions are left open. Firstly, can we outperform FCRS in terms of the availablity vs. repair bandwidth trade-off with a new architecture? Secondly, we saw that under the exact repair model, Cubic Codes minimize the repair bandwidth for FCRS with 3 complete clusters. Does this optimality result generalize to more than 3 clusters or the case of incomplete clusters? This question can be answered in the affirmative if at the Minimum Bandwidth Regenerating point for FCRS we have $\alpha = d\beta$, as this would imply Equation \eqref{eqn:theonlyrequirement}. While this may sound intuitively true, we do not have a proof for it. Finally, a more ambitious goal would be to design exact repair codes for FCRS for points other than MBR. Figure \ref{fig:comparisonfunctional} tells us that there is an interval near the MBR point at which FCRS outperforms the model in \cite{dimakis2010network} in the functional repair paradigm. By designing explicit codes for these inner points or establishing exact-repair converse bounds, one could try to prove or reject the hypothesis that within the same interval FCRS is superior in the exact repair model too.
\bibliographystyle{IEEEtran}
\bibliography{IEEEfull,fcrs}

\end{document}